\documentclass[useAMS,usenatbib,usegraphicx]{mn2e}

\usepackage{lscape}
\usepackage{color}
\usepackage[usenames,dvipsnames,svgnames,table]{xcolor}

\newcommand\aj{{AJ}}
\newcommand\araa{{ARA\&A}}
\newcommand\apj{{ApJ}}
\newcommand\apjl{{ApJ}}
\newcommand\apjs{{ApJS}}
\newcommand\apss{{Ap\&SS}}
\newcommand\aap{{A\&A}}
\newcommand\mnras{{MNRAS}}
\newcommand\pasp{{PASP}}
\newcommand\pasj{{PASJ}}

\title[Hidden population of AM CVn binaries in the SDSS]{A search for the hidden population of AM CVn binaries in the Sloan Digital Sky Survey}
\author[P. J. Carter et al.]{P. J. Carter,$^{1}$\thanks{E-mail: philip.carter@warwick.ac.uk} T. R. Marsh,$^{1}$ D. Steeghs,$^{1}$ P. J. Groot,$^{2}$ G. Nelemans,$^{2,3}$\newauthor D. Levitan,$^{4}$ A. Rau,$^{5}$ C. M. Copperwheat,$^{1}$ T. Kupfer$^{2}$ and G. H. A. Roelofs$^{6}$\\
$^{1}$Department of Physics, University of Warwick, Coventry CV4 7AL\\
$^{2}$Department of Astrophysics/IMAPP, Radboud University Nijmegen, PO Box 9010, 6500 GL Nijmegen, the Netherlands\\
$^{3}$Institute for Astronomy, KU Leuven, Celestijnenlaan 200D, 3001 Leuven, Belgium\\
$^{4}$Division of Physics, Mathematics, and Astronomy, California Institute of Technology, Pasadena, CA 91125, USA\\
$^{5}$Max--Planck Institute for Extraterrestrial Physics, Giessenbachstr. 1, Garching 85748, Germany\\
$^{6}$Harvard--Smithsonian Centre for Astrophysics, 60 Garden Street, Cambridge, MA 02138, USA\\
}

\begin{document}

\date{Accepted 2012 November 23}

\pagerange{\pageref{firstpage}--\pageref{lastpage}} \pubyear{2012}

\maketitle

\label{firstpage}

\begin{abstract}
We present the latest results from a spectroscopic survey designed to uncover the hidden population of AM Canum Venaticorum (AM CVn) binaries in the photometric database of the Sloan Digital Sky Survey (SDSS). We selected $\sim$2000 candidates based on their photometric colours, a relatively small sample which is expected to contain the majority of all AM CVn binaries in the SDSS (expected to be $\sim$50).

\textcolor{black}{We present two new candidate AM CVn binaries discovered using this strategy: SDSS J104325.08+563258.1 and SDSS J173047.59+554518.5. We also present spectra of 29 new cataclysmic variables, 23 DQ white dwarfs and 21 DZ white dwarfs discovered in this survey.}

The survey is now approximately 70 per cent complete, and the discovery of seven new AM CVn binaries indicates a lower space density than previously predicted. From the essentially complete $g \le$ 19 sample, we derive an observed space density of (5~$\pm$~3)~$\times$~10$^{-7}$~pc$^{-3}$; this is lower than previous estimates by a factor of 3.

\textcolor{black}{The sample has been cross-matched with the {\itshape GALEX} All-Sky Imaging Survey database, and with Data Release 9 of the UKIRT (United Kingdom Infrared Telescope) Infrared Deep Sky Survey (UKIDSS). The addition of UV photometry allows new colour cuts to be applied, reducing the size of our sample to $\sim$1100 objects. Optimising our followup should allow us to uncover the remaining AM CVn binaries present in the SDSS, providing the larger homogeneous sample required to more reliably estimate their space density.}
\end{abstract}

\begin{keywords}
accretion, accretion discs -- binaries: close -- stars: individual: SDSS J104325.08+563258.1, SDSS J173047.59+554518.5 -- novae, cataclysmic variables -- white dwarfs.
\end{keywords}

\section{Introduction}

The AM Canum Venaticorum (AM CVn) binaries are a class of ultracompact systems that consist of a white dwarf accreting helium rich material from a (semi-)degenerate donor. They are characterised by their short orbital periods, which range from 5 to $\sim$65 minutes, and an absence of hydrogen in their spectra. It is the degenerate nature of the mass donor that allows their periods to lie well below the believed minimum period spike ($\sim$80 min) of hydrogen rich cataclysmic variables (CVs; \citealt{1982ApJ...254..616R,2009MNRAS.397.2170G}). Only 33 members of the rare AM CVn binary class have been reported in the literature (the most recent by \citealt{2012MNRAS.........L}), 7 of which were discovered via systematic searches in the Sloan Digital Sky Survey (SDSS; \citealt{2000AJ....120.1579Y,2005MNRAS.361..487R,2005AJ....130.2230A,2008AJ....135.2108A}). A recent review of the AM CVn binaries is given by \citet{2010PASP..122.1133S}.

The observational features of an AM CVn binary are thought to depend on its orbital period, and it will pass through several distinct phases as the system evolves from period minimum towards longer orbital periods. In the shortest period systems ($P_{\rmn{orb}} < \sim $10 min) the accretion stream impacts directly onto the surface of the accretor, and no disc forms \citep{2002MNRAS.331L...7M,2010ApJ...711L.138R}. For orbital periods of $\sim$10 -- $\sim$20 minutes, the accretion disc is in a stable `high' state, and spectra are dominated by helium absorption from the optically thick disc \citep{1994MNRAS.271..910O}. In the long period systems ($P_{\rmn{orb}} > \sim$40 min) the disc is in a stable low state; these systems typically lack photometric variability, and their optical spectra are dominated by helium emission lines \citep{1995Ap&SS.225..249W,2001ApJ...552..679R}. The intermediate period systems (20 $\le P_{\rm orb}\le$ 40 min) have unstable discs, and their appearance varies between that of the high-state and the low-state systems, analogously to the hydrogen-rich dwarf novae \citep{1997PASJ...49...75T,2012A&A...544A..13K,2012MNRAS.419.2836R}.

The AM CVn binaries are of great interest for many aspects of astrophysics. These systems represent the end product of several finely-tuned evolutionary pathways \citep{1991ApJ...370..615I,2001A&A...368..939N,2003MNRAS.340.1214P}, and as such are of great interest for binary stellar evolution theory. They are also of particular interest due to their gravitational wave emission, being some of the strongest known sources that would be detected by future low-frequency, space-borne gravitational wave detectors, the brightest systems acting as calibrators for such an experiment \citep{2006CQGra..23S.809S,2006MNRAS.371.1231R,2009CQGra..26i4030N,2012ApJ...758..131N}. Determining an accurate space density will allow improved constraints to be placed on the expected gravitational wave signal. The AM CVn binaries are closely related to the double degenerate pathway to Type Ia supernovae, and may themselves contribute to the supernova population \citep{2005ASPC..334..387S,2009ApJ...699.2026R}. They may also harbour dynamical time scale helium fusion, producing rare sub-luminous SN Ia-like events (`SN.Ia'; \citealt{2007ApJ...662L..95B,2011MNRAS.411L..31B}). Establishing the space density is important for determining the stability of mass transfer in double white dwarf binaries \citep{2004MNRAS.350..113M}, which are potential progenitors for Type Ia supernovae, sdB stars and R CrB stars \citep{1984ApJ...277..355W}.

The first AM CVn binaries were discovered serendipitously in a number of different ways. \citealt{2001A&A...368..939N} conducted a population synthesis based on the two main proposed formation channels for AM CVn binaries, concluding that improved observations were required in order to learn more about the AM CVn population. The discovery of six AM CVn binaries in the SDSS spectroscopic database, via their helium emission dominated spectra \citep{2005MNRAS.361..487R,2005AJ....130.2230A,2008AJ....135.2108A}, provided the first relatively well defined, homogeneous sample, that allowed estimation of the local space density by calibrating the theoretical population models. This gave a value of 1--3~$\times$~10$^{-6}$ pc$^{-3}$ \citep{2007MNRAS.382..685R}. This was an order of magnitude lower than the expected space density at the time, derived from population synthesis (2~$\times$~10$^{-5}$ pc$^{-3}$, based on \citealt{2001A&A...368..939N}), but the accuracy was limited by the small sample size.

Since the known AM CVn binaries have been found to occupy a relatively sparsely-populated region of colour space, which has a low spectroscopic completeness in the SDSS database, we have initiated a dedicated spectroscopic survey designed to uncover the `hidden' population of AM CVn binaries in the SDSS photometry. Details of this programme, including the sample selection criteria, and the first new AM CVn binary, were presented by \citet{2009MNRAS.394..367R}; four further discoveries were described by \citet{2010ApJ...708..456R}.

The sample has since been extended to include new targets from SDSS Data Release 7. This increase in the survey area should lead to a corresponding increase in the size of the resulting AM CVn sample (which is desirable given the still small number of systems found), however, it also increases the observation time required to complete the programme. We therefore investigate the possibility of increasing our AM CVn detection efficiency using new colour--colour cuts based on photometry at longer and shorter wavelengths than the five SDSS bands provide. Here we present an update on the status of the programme, detail our findings so far, and propose improved selection criteria to increase our efficiency.

We describe the status of the survey and our observations in Section 2. In Section 3 we discuss cross-matching to the \textit{GALEX} All-Sky Imaging Survey and the UKIRT (United Kingdom Infrared Telescope) Infrared Deep Sky Survey. In Section 4 we discuss our identification procedure, and present our CV, DQ white dwarf and DZ white dwarf samples. In Section 5 we detail our results, present two new candidate AM CVn binaries, describe new color cuts based on \textit{GALEX} photometry, and report our revised value for the space density. We discuss these results in Section 6.

\section{Survey observations}

\begin{figure}
 \centering
 \includegraphics[width=0.48\textwidth]{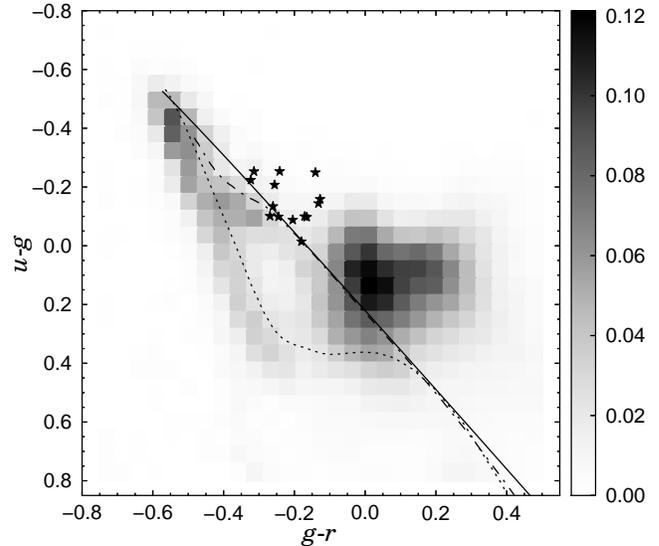}
 \caption{Completeness of SDSS DR9 spectroscopic follow-up as a function of colour, $u-g$ and $g-r$, to a limiting magnitude $g=20.5$ (dereddened). The long period SDSS AM CVn binaries are indicated by star symbols. The solid line marks the blackbody cooling track, the dotted and dot-dashed lines indicate model cooling sequences for DA and DB white dwarfs. \label{f:sloancompleteness}} 
\end{figure}
Fig. \ref{f:sloancompleteness} shows the spectroscopic completeness of the SDSS Data Release 9 as a function of $u-g$ and $g-r$ colour, to a limiting $g$-band magnitude of 20.5. Also plotted is the blackbody cooling track, and Bergeron model colours of the cooling sequences of hydrogen (DA) and helium (DB) atmosphere white dwarfs (log $g = 8.0$; \citealt{2006AJ....132.1221H,2006ApJ...651L.137K,2011ApJ...730..128T,2011ApJ...737...28B}). The known AM CVn binaries lie in an area of relatively low spectroscopic coverage, largely offset from the DB white dwarf track. The areas to the lower left and right are targeted for DA white dwarfs and quasars.

As shown by \citet{2009MNRAS.394..367R}, AM CVn binaries with a He~\textsc{i} 5875 equivalent width $\ge$30 \AA{} are easily detectable at a signal-to-noise ratio of 10, and resolutions as low as $R = 300$. Even systems with weaker lines should still show some sign of helium emission that would be confirmed by a second, higher quality spectrum. The possibility that there exist some systems with much weaker lines was discussed by \citet{2009MNRAS.394..367R}. They concluded that the majority of the population should be detectable.

\begin{figure}
 \centering
 \includegraphics[width=0.48\textwidth]{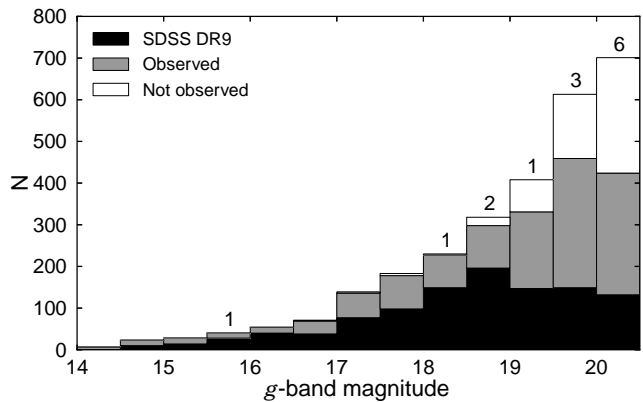}
 \caption{Spectroscopic completeness of our sample. We show objects meeting our selection criteria for which there are Sloan spectra (black), targets for which we have taken our own spectra (grey), and targets yet to be observed (empty). Numbers above bins indicate the number of currently known emission line AM CVn binaries in our spectroscopy or the SDSS DR9 spectroscopic database. \label{f:completeness}}
\end{figure}

Low-resolution, low signal-to-noise ratio spectra of objects in our colour-selected sample have been obtained with several different instrument setups, using a variety of low and intermediate resolution spectrographs. To date, spectroscopic observations of 1403 of the 1947 candidates have been completed (see Fig. \ref{f:completeness}).
\begin{table*}
\begin{minipage}{172mm}
 \centering
 \caption{Summary of our observing campaign and the instrument setups used. The wavelength range and resolution achieved with each instrument are listed, as well as the number of candidates observed during each programme. Several of our candidates from SDSS DR7 have been observed by the SDSS as part of more recent data releases, those we have not observed previously ourselves are also listed here.}
 \label{t:obslog}
 \begin{tabular}{l l r r r}
  \hline
  Dates 			& Telescope/Instrument 	& Number of candidates observed & Wavelength range (\AA) 	& R at 5875\AA \\
  \hline
  2008 Jan 09 -- 2009 Apr 30 	& Tillinghast/FAST 	& 225 				& 3500 -- 7400 			& 810 \\
  2008 Feb 25 -- 2008 Mar 02 	& INT/IDS 		& 117 				& 3500 -- 8500			& 630 \\
  2008 Feb 27 			& Keck-I/LRIS 		& 6 				& 3400 -- 5700, 6750 -- 9000	& 730 \\
  2008 June 06 -- 2008 Sep 20 	& VLT/FORS1 		& 18 				& 3600 -- 8600			& 340 \\
  2008 June 03 -- 2009 Apr 25 	& Hale/DBSP	 	& 143 				& 3400 -- 8000			& 540 \\
  2008 Dec 25 -- 2008 Dec 28 	& WHT/ISIS 		& 68 				& 3200 -- 8200			& 1160 \\
  2009 Feb 26 -- 2009 Apr 22 	& Gemini-South/GMOS 	& 14 				& 3900 -- 6700			& 640 \\
  2009 July 21 -- 2009 Aug 24 	& Gemini-North/GMOS 	& 15 				& 3900 -- 6700			& 640 \\
  2009 Mar 17 -- 2009 Mar 23 	& INT/IDS 		& 95 				& 3500 -- 8500			& 630 \\
  2009 May 24 -- 2009 May 28 	& NOT/ALFOSC 		& 131 				& 3800 -- 9000			& 180 \\
  2009 June 18 -- 2009 June 24 	& WHT/ACAM 		& 205 				& 3800 -- 9200			& 300 \\
  2009 Oct 09 -- 2009 Oct 15 	& NTT/EFOSC 		& 59 				& 3800 -- 8000			& 340 \\
  2009 Nov 08 -- 2009 Nov 14 	& WHT/ACAM 		& 130 				& 3800 -- 9200			& 300 \\
  2010 Aug 19 -- 2010 Aug 23 	& WHT/ACAM 		& 88 				& 3800 -- 9200			& 300 \\
  2010 Nov 01 -- 2010 Nov 06 	& WHT/ACAM 		& 5 				& 3800 -- 9200			& 300 \\
  2011 Feb 02 -- 2011 Feb 06 	& NOT/ALFOSC 		& 55 				& 4000 -- 9000			& 180 \\
  2012 May 22 -- 2012 May 23	& SOAR/Goodman		& 5				& 3800 -- 7000			& 860 \\
  2012 June 13			& Keck-I/LRIS		& 1				& 3100 -- 10000			& 2100 \\
  2012 July 13			& WHT/ISIS		& 4				& 3800 -- 5200, 5600 -- 7100	& 2200 \\
  2008 July -- 2009 June 	& SDSS DR8		& 17				& 3800 -- 9200			& 1800 \\
  2009 June -- 2011 July 	& SDSS DR9		& 11				& 3650 -- 10400			& 1800 \\
  \hline
 \end{tabular}
\end{minipage}
\end{table*}
These data were obtained using a number of telescopes: the Isaac Newton Telescope (INT), William Herschel Telescope (WHT), and Nordic Optical Telescope (NOT), on the island of La Palma; the 1.5-m Tillinghast telescope at the Fred Lawrence Whipple Observatory, Mt. Hopkins, Arizona; the Very Large Telescope (VLT), at Paranal; the 200-inch Hale telescope at the Palomar Observatory; both Gemini telescopes, South in Chile and North on Hawaii; and the New Technology Telescope (NTT), at La Silla. The largest telescopes, VLT and Gemini, have been used to observe the faintest part of the sample, $g$-band magnitude $> 20$. A log of our observations, listing the instrument setups used and the numbers of candidates observed, is given in Table \ref{t:obslog}.

Instrument setups were chosen such that all spectra cover the 4000--7000 \AA{} wavelength range that includes the predominant lines of both helium and hydrogen, allowing identification of AM CVn binaries from their strong helium emission, and lack of hydrogen.

The spectra obtained with the FAST spectrograph were reduced using the spectral extraction pipeline provided by the observatory. This pipeline is based on standard \textsc{IRAF} routines, see \citet{1997ASPC..125..140T}. All other data were reduced using optimal extraction as implemented in the \textsc{Pamela}\footnote{\textsc{Pamela} is included in the \textsc{Starlink} distribution `Hawaiki' and later releases. The \textsc{Starlink} Software Group homepage can be found at http://starlink.jach.hawaii.edu/starlink.} code \citep{1989PASP..101.1032M}, and the \textsc{Starlink} packages \textsc{Kappa}, \textsc{Figaro} and \textsc{Convert}. Wavelength calibration was obtained from various arc lamp exposures taken each night. Flux calibration was achieved with various standard stars observed at the beginning or end of each night. The flux calibration is not absolute, and approximately one third of the spectra do not have flux calibration.

Fig. \ref{f:completeness} shows the current spectroscopic completeness of our sample. The target list at the bright end of the distribution has been effectively completed, allowing us to draw some preliminary conclusions. We estimate the number of systems in our sample, from the numbers we have found so far, by multiplying the numbers of AM CVn binaries found in each magnitude bin in Fig. \ref{f:completeness} with the ratio of the total to observed number per bin. This suggests that there should be at least 5 more systems yet to be discovered in our sample, however, we note that with such small numbers there is a large uncertainty.

\section{Cross-matching}

\subsection{\textit{GALEX}}

The \textit{GALEX} satellite has conducted a series of imaging and spectroscopic surveys, including the first UV all sky survey \citep{2005ApJ...619L...1M}. This consists of imaging in two broad bands, near-ultraviolet (NUV, 1770--2730 \AA) and far-ultraviolet (FUV, 1350--1780 \AA), of $\sim$26,000 square degrees of sky, to a depth of 20.5 mag (\textit{GALEX} uses the AB photometric system). This is complemented by medium and deep imaging surveys with greater depths, covering smaller areas. All candidates from the SDSS DR7 colour selection were matched against the \textit{GALEX} Data Release 6 catalogue, taking the closest neighbour within 2 arcsec. This matching radius was chosen to ensure good coverage of the expected distribution of offsets between \textit{GALEX} and SDSS astrometry, without introducing a significant number of false or multiple matches \citep{2007ApJS..173..682M}.

We do not use the predefined \textit{GALEX}-SDSS cross-match described by \citet{2009ApJ...694.1281B}, as we prefer the smaller FUV-NUV match radius used for the standard catalogs, in order to minimize false matches \citep{2007ApJS..173..682M}. We also prefer to have greater freedom in the choice of \textit{GALEX}-SDSS matching radius. We expect that the results should be identical in most cases.

We estimate the false match probability by repeating the search of the \textit{GALEX} database after applying a random 10 arcsec offset to the coordinates of each SDSS source for which a \textit{GALEX} counterpart has been found. The number of matched coordinates divided by the total number of coordinates, gives an estimate of the false match rate; we derive a value of approximately 3.5 per cent.

A total of 1622 of the 1947 SDSS DR7 objects have at least one measured UV magnitude, 1590 of these have a NUV detection, and 1138 are detected in both \textit{GALEX} bands.

\subsection{UKIDSS}

The SDSS DR7 sample was also matched to the UKIRT (United Kingdom Infrared Telescope) Infrared Deep Sky Survey (UKIDSS) DR9 catalogue.
When complete, the UKIDSS Large Area Survey will cover 4028 square degrees of sky, with imaging in four broad-band filters, $Y$, $J$, $H$ and $K$, to a $Y$-band limiting magnitude of 20.3 \citep{2006MNRAS.372.1227D}. A matching radius of 1.0 arcsec was used for the UKIDSS catalogue (which has a greater astrometric accuracy than \textit{GALEX}), this covers the majority of the offset distribution.

Again we estimate the false match rate by offsetting the coordinates of the matched sample and repeating the search. We find a false match probability of $\sim$1 per cent for the UKIDSS. Of our 1947 SDSS targets, only 516 are in the area that has been covered by UKIDSS, and 398 have a detection in at least one filter.

\section{Spectroscopic identification}

\begin{figure*}
 \includegraphics[width=1.0\textwidth]{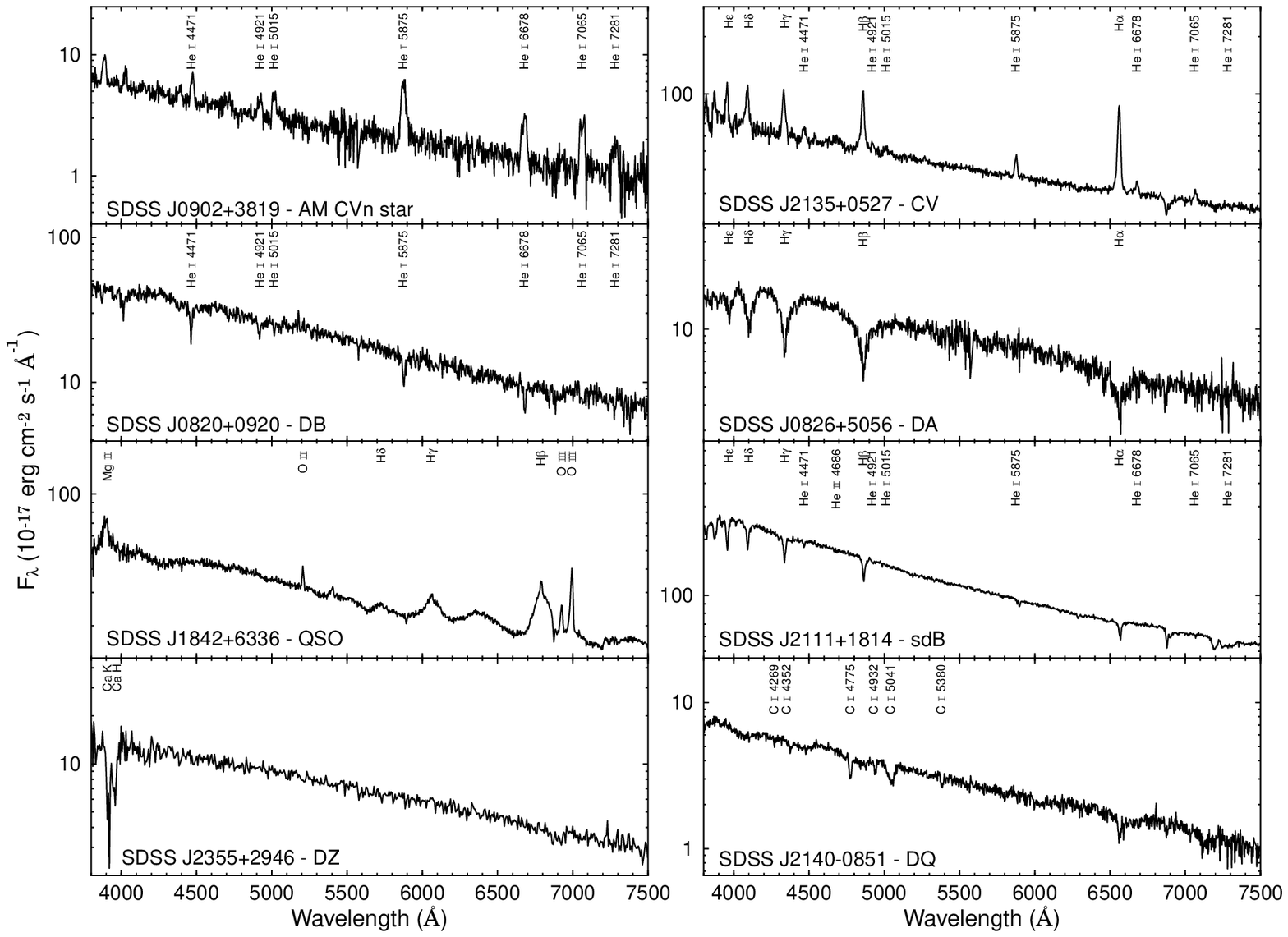}
 \caption{Identification spectra of select objects from our sample, demonstrating the appearance of the main object classes. The features that identify each object as a member of its class are labelled. The emission lines of helium and hydrogen present in the spectra of AM CVn binaries and CVs produce a strong contrast between them and the white dwarf contaminants, even at low S/N. The extremely broad, redshifted emission lines seen in quasar spectra, distinguish them from the other object types. \label{f:examplespectra}}
\end{figure*}

We assign a spectroscopic classification to each object based on visual inspection of its spectrum.
\begin{table}
 \centering
 \caption{Numbers of main object types identified in our sample. Objects classified as DA, DB, DQ and DZ white dwarfs have spectra dominated by hydrogen, helium, carbon and metal lines respectively; DC white dwarfs are those that exhibit a continuum spectrum. The subdwarf classification includes objects identified as sdB, sdOB or sdO.}
 \label{t:class}
 \begin{tabular}{l r}
  \hline
  Class 	& Number \\
  \hline
  AM CVn 	& 7 \\
  CV		& 29 \\ 
  quasar        & 109 \\
  galaxy	& 2 \\
  white dwarfs:	& \\
  \ DA		& 120 \\
  \ DB		& 427 \\
  \ DQ		& 23 \\
  \ DZ		& 21 \\
  \ DC		& 73 \\
  WD+dM  	& 1 \\
  subdwarf	& 184 \\
  unknown	& 407 \\
   & \\
  Total		& 1403 \\
  \hline
 \end{tabular}
\end{table}
Table \ref{t:class} gives the numbers of Cataclysmic Variables (CVs; see \citealt{1995CAS....28.....W} for a review), AM CVn binaries, white dwarfs, subdwarfs (see \citealt{2009ARA&A..47..211H} for a recent review) and other objects identified in our sample. The white dwarfs are divided into several subclasses, DA white dwarfs are those with hydrogen atmospheres, DB those with helium atmospheres, DQ those with carbon dominated spectra, DZ those with spectra dominated by metal lines; and DC those with a continuum spectrum (see \citealt{1993PASP..105..761W} for an overview of white dwarf spectra). Fig. \ref{f:examplespectra} shows example spectra of these main classes, with the features that identify them labelled. The AM CVn binaries, CVs and quasars are easily identified by their emission lines; the systems that show only absorption in their spectra often have less certain classifications. A full list of candidates with classifications, coordinates, FUV, NUV and $u$, $g$, $r$, $i$, $z$ magnitudes is given in Table \ref{t:catalogue}\footnote{\textcolor{black}{The full spectroscopic catalogue is available in electronic form at the CDS
(http://cdsweb.u-strasbg.fr/).}}.

\begin{figure*}
 \centering
 \includegraphics[width=0.49\textwidth]{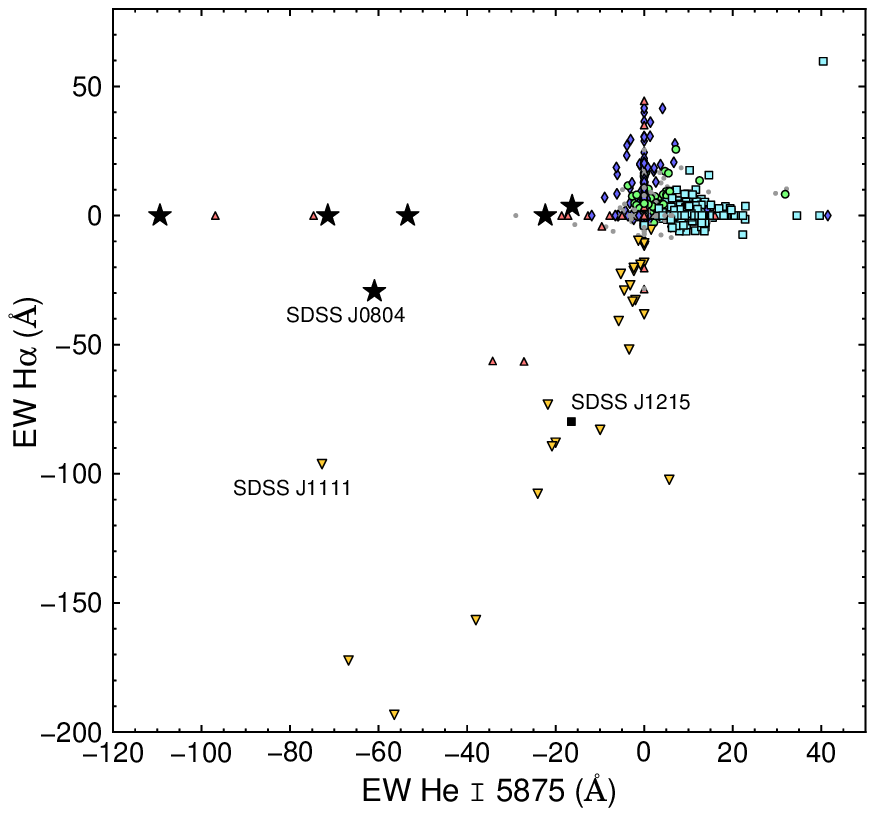}
 \includegraphics[width=0.49\textwidth]{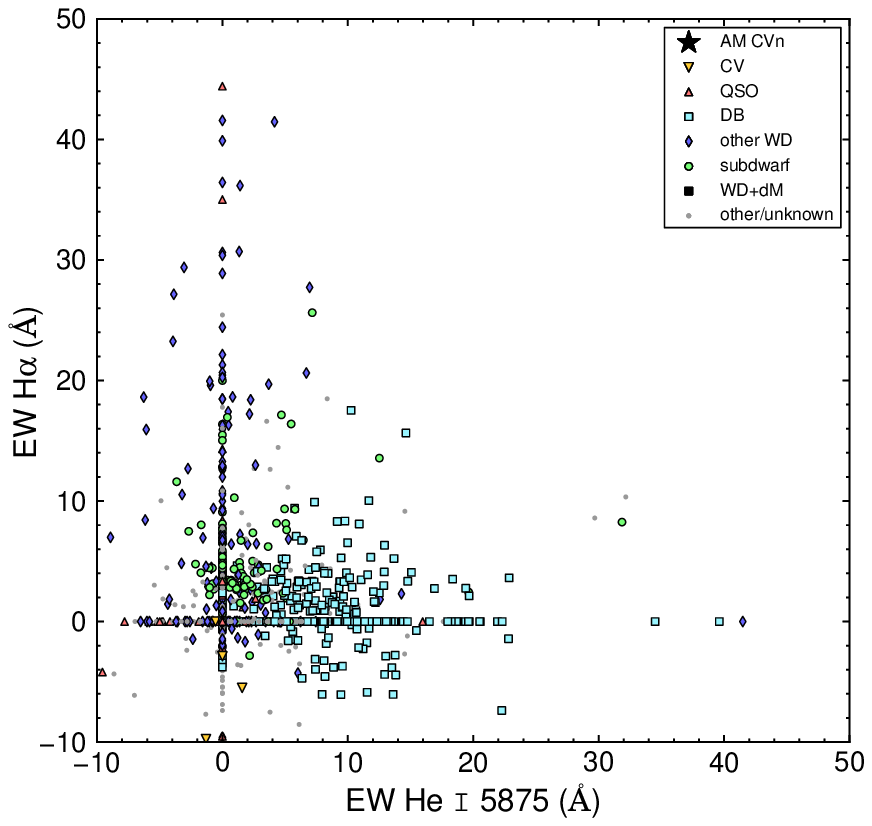}
 \caption{Equivalent width of He~\textsc{i} 5875 vs H$\alpha$ for our sample, the right hand panel shows the crowded area of the diagram, occupied mainly by white dwarfs, in more detail.
 The black stars and orange inverted triangles represent AM CVn binaries and CVs. Quasars, DB white dwarfs, other types of white dwarf, subdwarfs, and WD + M dwarf systems are represented by red triangles, cyan squares, blue diamonds, green circles and black squares respectively. Grey dots indicate those candidates that could not be classified.
 Equivalent widths have been calculated using a Gaussian plus sloped continuum fit; the large values for some unclassified objects are largely the result of noise. Objects appear in the expected regions with allowance for some scatter caused by noise. Note that He~\textsc{ii} 6559 emission would not be distinguished from H$\alpha$ by the fitting algorithm. Several quasars have apparently valid equivalent widths for some lines, and appear randomly distributed, but are distinguishable using other spectral features. \label{f:ew5875Ha}} 
\end{figure*}

Fig. \ref{f:ew5875Ha} shows the He~\textsc{i} 5875 against H$\alpha$ equivalent width distribution of the spectra of our sample (here emission lines correctly have negative equivalent widths, this is ignored for simplicity elsewhere in the text). These were calculated using a Gaussian fit to the normalised spectrum, with fit constraints chosen to be consistent with the line fitting algorithm used by the SDSS. If the fit failed, the offset from the expected central wavelength was larger than 20 \AA{} or the Gaussian dispersion was measured to be less than 0.5 \AA{} or greater than 100 \AA{}, the result was rejected and a value of zero taken. Objects appear in the expected regions of the diagram, with some scatter caused by noise.

The AM CVn binaries (stars) lie close to the line EW(H$\alpha$) = 0 \AA{}, with negative He~\textsc{i} 5875 equivalent widths. SDSS J0804+1616 has significant He~\textsc{ii} 6559 emission that causes the apparent large equivalent width for H$\alpha$. The CVs (orange inverted triangles) are all found below the line EW(H$\alpha$) = EW(5875), and (with the exception of SDSS J203311.78+134954.1, probably due to interstellar absorption) to the left of EW(5875) = 0 \AA{}. As would be expected, the CVs identified all have hydrogen emission, which is always stronger than helium emission. SDSS J121534.77+025726.6, identified as a white dwarf + M dwarf binary, is also found in this part of the diagram due to the hydrogen emission in its spectrum (most likely due to irradiation or chromospheric activity). Several quasars have apparently valid equivalent widths for some lines, and are scattered over the diagram; these are easily distinguished by visual inspection of their spectra, and also by their colour, which could be useful to distinguish those with low signal-to-noise spectra falling in the AM CVn region of Fig. \ref{f:ew5875Ha}. That there are no significant large deviations from the expected position for any object type, suggests that our classification by visual inspection was successful.

\textcolor{black}{There are 407 spectra which have no clear identifiable features, so no likely classification has been assigned. These objects are generally fainter, resulting in lower signal-to-noise ratio spectra, which makes weak features more difficult to discern. The majority of these objects are likely white dwarfs with weak lines or DC white dwarfs; we do not assign this classification as we do not want our assumptions as to their nature to affect our understanding of the colour space. With the exception of 18 candidates, none of these objects have helium emission with an equivalent width greater than 3 \AA{} as the dominant feature in their spectra, ruling them out as possible AM CVn binaries. The measured equivalent widths for several of these exceptions that have possible helium emission, have very large errors, and the measurements are rejected on this basis. Fourteen of these unclassified objects are not consistent with having an He~\textsc{i} equivalent width of 0 \AA{}; although the apparent emission would be classified as noise by visual inspection, they cannot be ruled out as AM CVn binaries and should be re-observed to obtain higher signal-to-noise spectra.}

The AM CVn binaries discovered in the SDSS spectroscopic database, and via this survey, all have equivalent width $> 13$ \AA{} for one of the 4686 \AA{}  or 5875 \AA{}  helium emission lines. This indicates our ability to identify candidate AM CVn binaries matches that of previous searches of the SDSS spectroscopic database \citep{2005AJ....130.2230A,2005MNRAS.361..487R,2008AJ....135.2108A}; and we can be confident that we would be able to detect the AM CVn binaries expected from predictions \citep{2007MNRAS.382..685R,2009MNRAS.394..367R}.

\textcolor{black}{It must also be noted that high-state AM CVn binaries -- those with short orbital periods and those in outburst -- typically show absorption, rather than emission, in their spectra, due to the optically thick accretion disc. The spectra of these systems look very similar to those of DB white dwarfs, AM CVn itself was originally identified as such \citep{1957ApJ...126...14G}, there are however, some differences in the line strengths that allow them to be distinguished with high-quality spectra (e.g. \citealt{1975ApJ...200L..23R,2011ApJ...726...92F}). At the low spectral resolutions and signal-to-noise ratios employed in this survey, these high-state systems are essentially indistinguishable from DB white dwarfs, or possibly DC white dwarfs (e.g. \citealt{1987MNRAS.227..347O}).}

The short period AM CVn binaries are expected to be far less numerous at the high Galactic latitudes of the SDSS, as the evolution from turn-on of mass transfer to $P_{\rmn{orb}} > 30$ min is relatively rapid \citep{2001A&A...368..939N}. Using a simple model for the evolution of a white dwarf channel system \citep{2001A&A...368..939N}, we find that an AM CVn binary with typical values for the mass at period minimum (following \citealt{2011ApJ...739...68L}), has an orbital period below 30 minutes for 5.8 per cent of its life between period minimum and $P_{\rmn{orb}}$ = 60 minutes. The high state accretion discs in the short period systems also make them much brighter, resulting in increased numbers in a magnitude limited sample. With an estimate of both the relative numbers and their absolute magnitude as a function of period, it is possible to calibrate the AM CVn space density based only on the emission line systems \citep{2007MNRAS.382..685R}.

\subsection{Cataclysmic variables}

\begin{figure*}
 \includegraphics[width=1.0\textwidth]{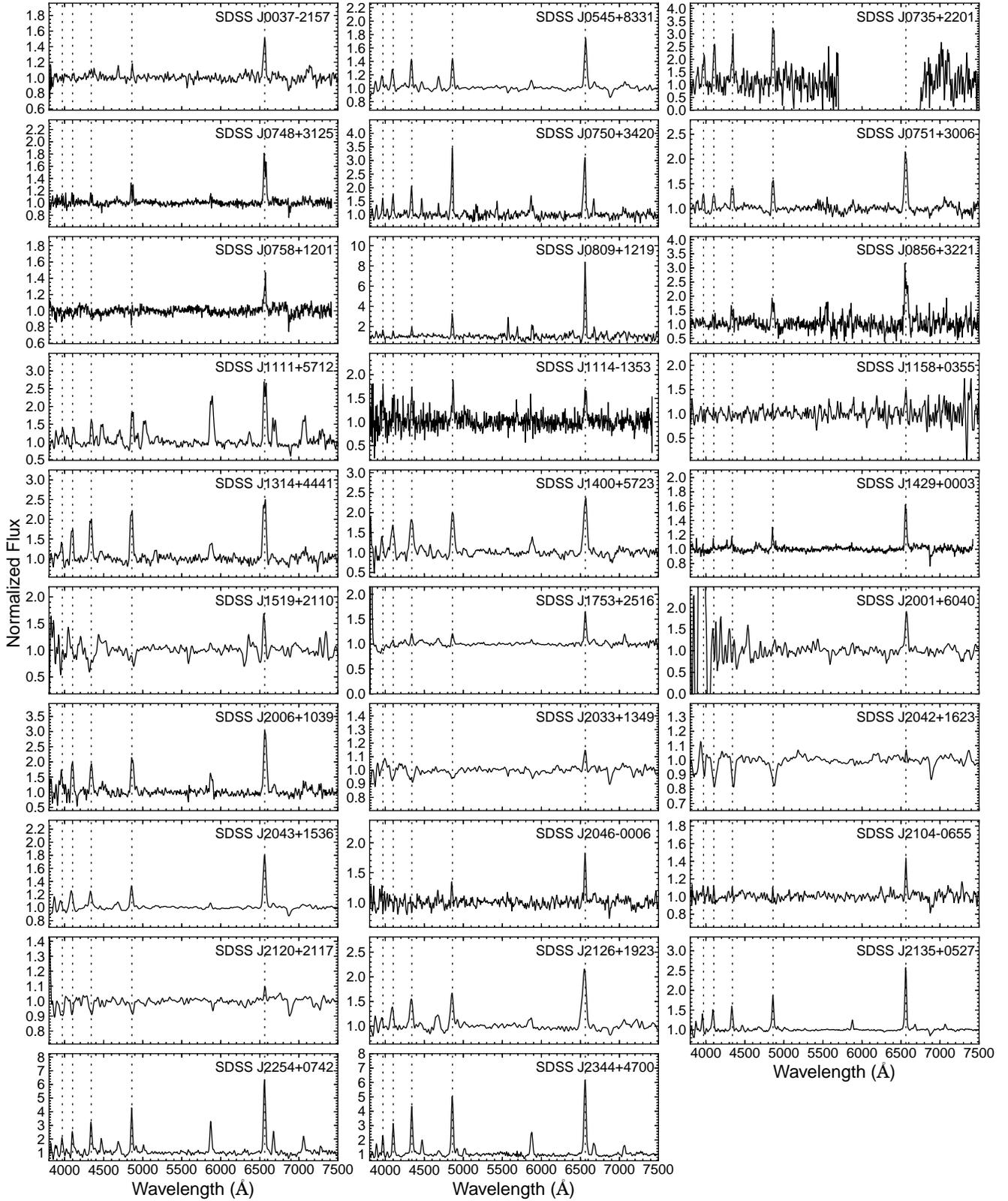}
 \caption{Identification spectra of the CVs discovered in our sample, displayed with 2 pixel Gaussian smoothing. Dotted lines indicate the wavelengths of the strongest hydrogen lines.\label{f:CVspectra}}
\end{figure*}

Fig. \ref{f:CVspectra} shows the spectra of the CVs identified in our sample via their hydrogen emission. The range of line strengths and ratios seen in Fig. \ref{f:ew5875Ha} are clear. The broad double-peaked line profile characteristic of an accretion disc is evident in several objects.

SDSS~J074859.54+312512.7,~SDSS~J075107.51+300628.5 and SDSS~J131432.11+444138.8, were previously identified as dwarf nova candidates by \citet{2010MNRAS.402..436W} based on their variability. The spectra presented here confirm these systems as CVs.
SDSS J225417.54+074227.3 (USNO-A2.0 0975-21112378) was previously identified as a QSO candidate by \citet{2007ApJ...664...53A}.

\begin{table*}
\begin{minipage}{108mm}
\centering
\caption{He~\textsc{ii} 4686 and H$\beta$ emission line equivalent widths and line ratios for CVs with strong He~\textsc{ii} emission.}
\label{t:4686ratio}
\begin{tabular}{l r r r}
\hline
Name				& EW He~\textsc{ii} 4686 (\AA)	& EW H$\beta$ (\AA)	& Line ratio \\
\hline
SDSS J003719.29$-$215714.4	& $-$3.3 $\pm$ 0.1		& $-$3.8 $\pm$ 0.1	& 0.87 $\pm$ 0.03 \\
SDSS J111126.83+571238.6	& $-$19.0 $\pm$ 2.0		& $-$34.0 $\pm$ 2.0	& 0.56 $\pm$ 0.07 \\
SDSS J175320.59+251649.1	& $-$6.0 $\pm$ 0.5 		& $-$6.9 $\pm$ 0.5	& 0.87 $\pm$ 0.10 \\
SDSS J204643.30-000630.2	& $-$3.8 $\pm$ 0.3 		& $-$5.9 $\pm$ 0.3	& 0.64 $\pm$ 0.06 \\
SDSS J212617.62+192320.1	& $-$18.0 $\pm$ 2.0 		& $-$33.0 $\pm$ 2.0	& 0.55 $\pm$ 0.07 \\
\hline
\end{tabular}
\end{minipage}
\end{table*}
We compute the He~\textsc{ii}~4686/H$\beta$ line ratios for the CVs with significant He~\textsc{ii} emission, in order to identify possible magnetic or nova-like systems. The five systems with an equivalent width ratio greater than 0.5 are listed in Table \ref{t:4686ratio}. The He~\textsc{ii} 4686 equivalent widths measured for SDSS J175320.59+251649.1 and SDSS J212617.62+192320.1 may be affected by blending with nearby lines, but are still large compared with most CVs. The weak lines exhibited by SDSS J003719.29$-$215714.4 and SDSS J175320.59+251649.1, combined with the strong He~\textsc{ii} suggest they may be high accretion rate nova-like systems. SDSS J111126.83+571238.6 also shows unusually strong He~\textsc{i} emission and is discussed further below.

\begin{figure}
\centering
\includegraphics[width=0.48\textwidth]{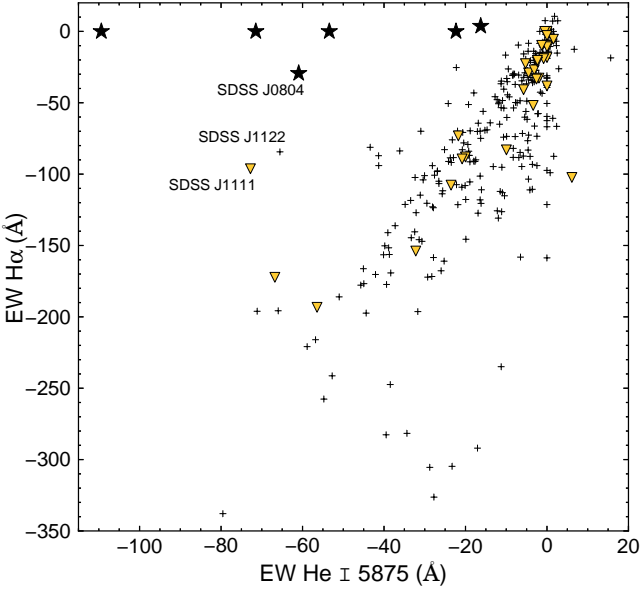}
\caption{\textcolor{black}{Equivalent width of He~\textsc{i} 5875 vs H$\alpha$ for our CV sample, and the Sloan CV population. Symbols have the same meanings as in Fig. \ref{f:ew5875Ha}, and the crosses represent the Sloan CVs.} \label{f:ewSDSSCVs}}
\end{figure}

Fig. \ref{f:ewSDSSCVs} shows the equivalent width distribution of our CV sample from Fig. \ref{f:ew5875Ha}, with the Sloan CV sample \citep{2011AJ....142..181S} added for comparison.
The distributions of the two CV populations look similar. The strongest H$\alpha$ emitters fall outside our colour box, and none of the CVs in our sample have H$\alpha$ emission as strong as the strongest Sloan CVs. There are two obvious outliers, SDSS J111126.83+571238.6 (SBSS 1108+574) from our sample and SDSS J112253.3$-$111037.6 (CSS100603:112253$-$111037, \citealt{2012MNRAS.425.2548B}) from the Sloan population, that have much stronger helium emission relative to hydrogen than the majority of CVs (this is also clear from the spectrum of SDSS J1111+5712 shown in Fig. \ref{f:CVspectra}). These may represent hybrid CV -- AM CVn binaries, and may be AM CVn binaries forming via the evolved CV formation channel \citep{2003MNRAS.340.1214P,2012MNRAS.425.2548B}.

\subsection{DQ white dwarfs}

\begin{figure*}
 \includegraphics[width=1.0\textwidth]{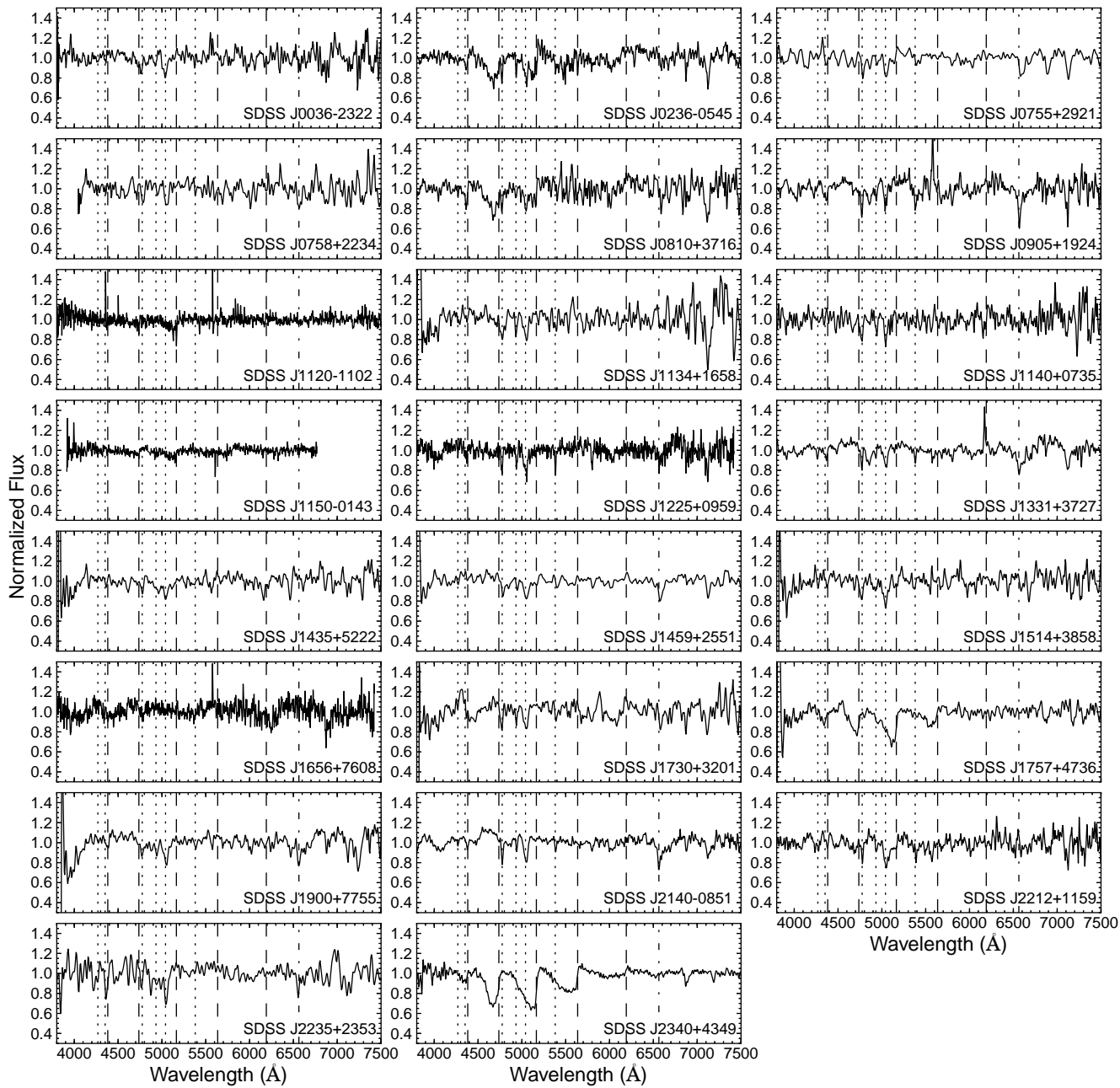}
 \caption{Identification spectra of the DQ white dwarfs discovered in our sample, displayed with 2 pixel Gaussian smoothing. Dotted lines indicate the wavelengths of the strongest lines of neutral carbon, dashed lines indicate the C$_{2}$ Swan band heads. The dot-dashed line indicates the wavelength of H$\alpha$, which is present in several objects. SDSS J1120$-$1102 was observed as part of SDSS DR8, and has not been observed separately by us.\label{f:DQspectra}}
\end{figure*}

Fig. \ref{f:DQspectra} shows the spectra of the DQ white dwarfs identified in our sample. These objects have been identified as DQs due to the presence of either neutral carbon lines or molecular C$_{2}$ Swan bands in their spectra. A number also show weak H$\alpha$ absorption and are classified as DQA.

The expected correlation of the transition from atomic to molecular carbon features with colour (or temperature), seen by \citet{2003AJ....126.1023H} in SDSS DQ white dwarfs, is also seen in our DQ sample.

\subsection{DZ white dwarfs}

\begin{figure*}
 \includegraphics[width=1.0\textwidth]{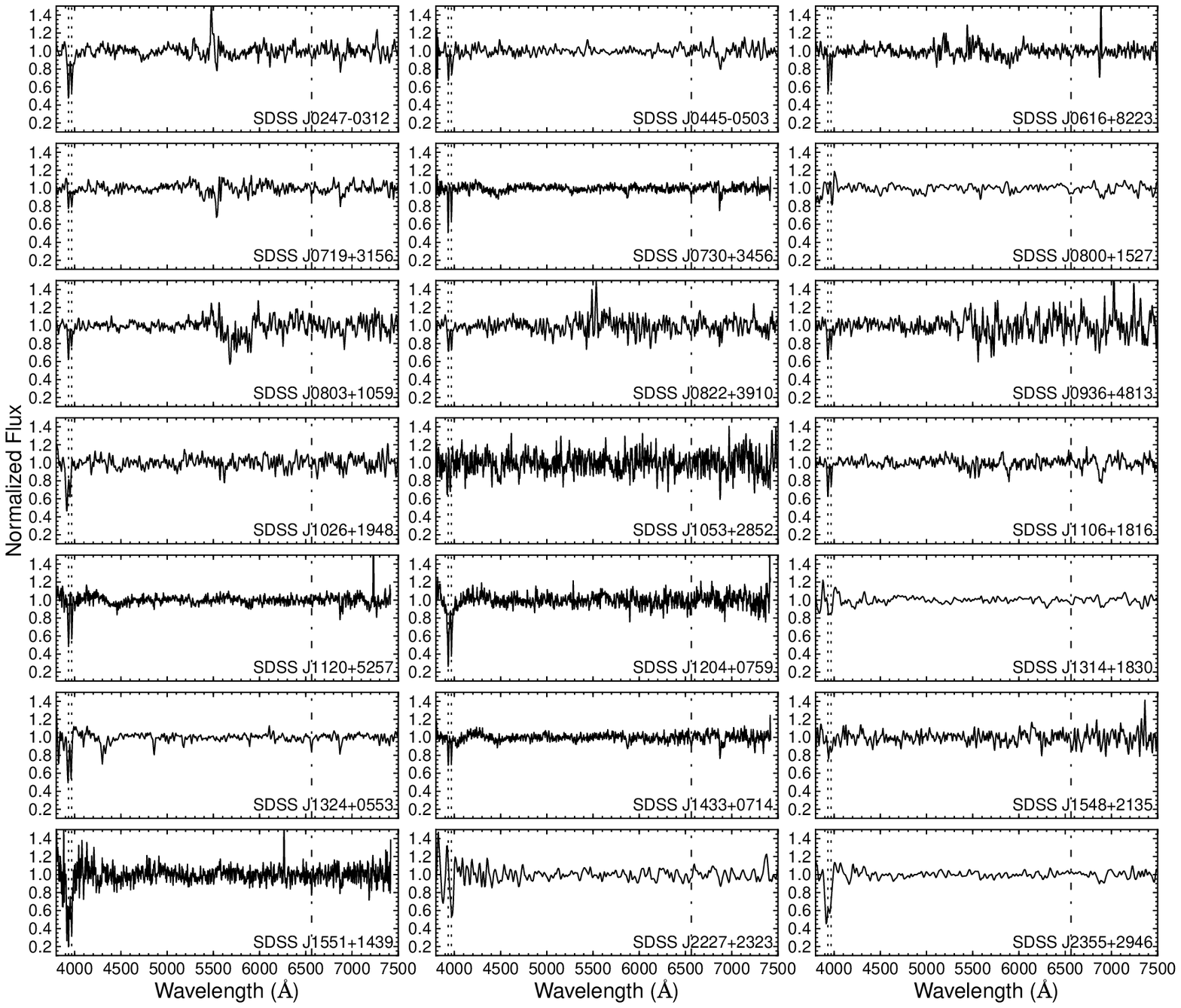}
 \caption{Identification spectra of the DZ white dwarfs discovered in our sample, displayed with 2 pixel Gaussian smoothing. Dotted lines indicate the wavelengths of the calcium H and K lines. The dot-dashed line indicates the wavelength of H$\alpha$, which is seen in several objects.\label{f:DZspectra}}
\end{figure*}

Fig. \ref{f:DZspectra} shows the spectra of the DZ white dwarfs identified in our sample. These objects have been identified as DZs due to the presence of the Ca \textsc{ii} H and K lines in their spectra. A number also show weak H$\alpha$ absorption and are classified as DZA.

SDSS~J105338.16+285245.6 (USNO-A2.0~1125-06201629) was previously identified as a QSO candidate by \citet{2007ApJ...664...53A}.

All the DZs in our sample show calcium absorption with no detection of other metals, similarly to the majority of SDSS DZ white dwarfs \citep{2003AJ....126.1023H}. They are all cool in ($u-g$, $g-r$), with the exception of SDSS J132430.43+055316.0, which also shows significant hydrogen in its spectrum.

\section{Results}

\subsection{Identification spectra of two new AM CVn binaries}
\label{sec:J1730}

\begin{figure*}
 \includegraphics[width=1.0\textwidth]{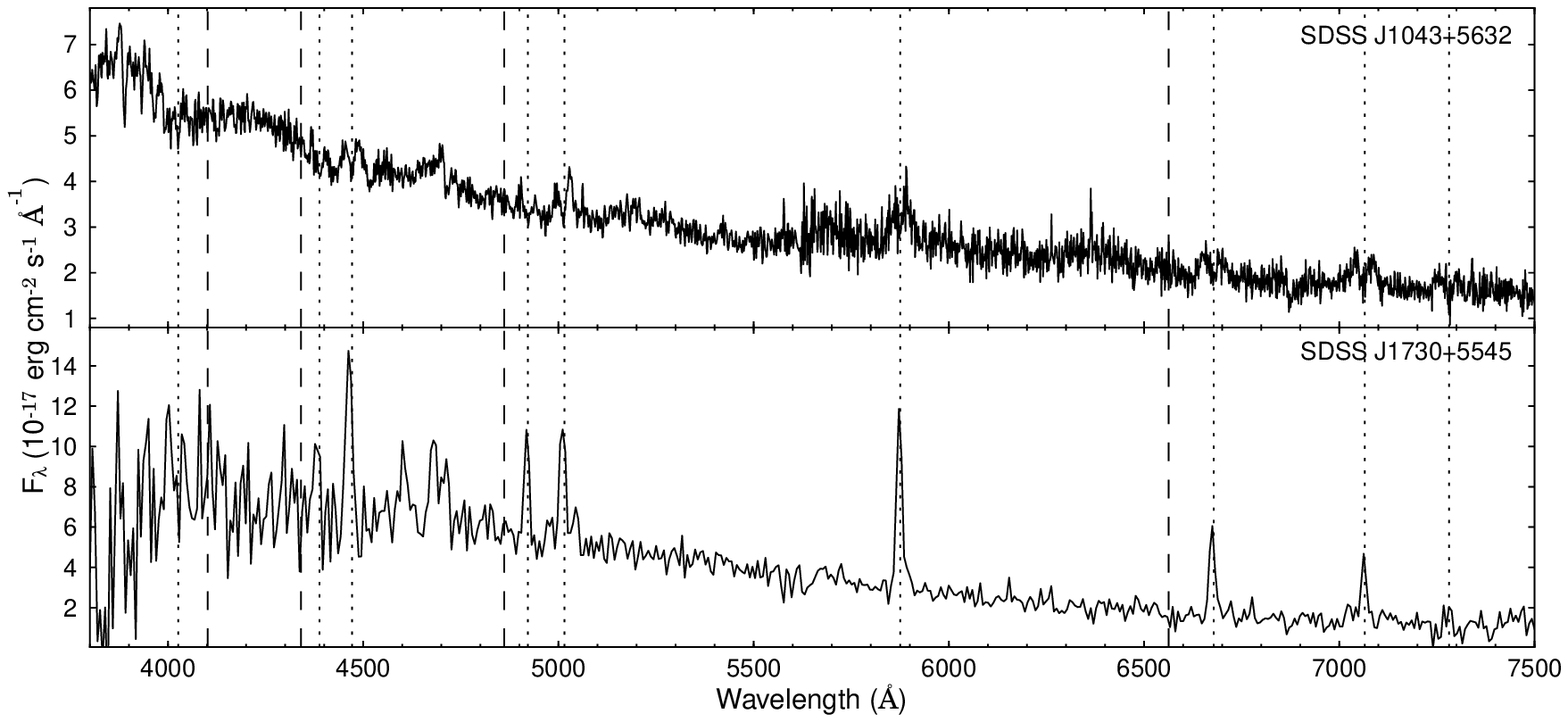}
 \caption{Identification spectra of SDSS J1043 (taken with LRIS on Keck I on 2012 June 13) and SDSS J1730 (taken with ACAM on the WHT on 2009 November 14). The wavelengths of the strongest He~\textsc{i} and hydrogen lines are indicated by the dotted and dashed lines. The lack of hydrogen features in the spectra is clear. SDSS J1043 shows clear double-peaked line profiles, in contrast to the narrow lines displayed by SDSS J1730, this is likely due to a lower inclination of SDSS J1730. SDSS J1043 also shows absorption wings around the He~\textsc{i} 4026 and 4471 \AA{} lines, thought to be due to the accretor (e.g.\ \citealt{2004RMxAC..20..254R}).\label{f:1730}}
\end{figure*}

The extracted spectra of SDSS J104325.08+563258.1 (hereafter SDSS J1043) and SDSS J173047.59+554518.5 (hereafter SDSS J1730) are shown in Fig. \ref{f:1730}.
The helium emission lines at 4471, 5875, 6678 and 7065 \AA{} are clearly detected, and no hydrogen is present. This suggests that SDSS J1043 and SDSS J1730 are likely new members of the AM CVn class, although further observations are required to confirm their ultra-compact binary nature and determine their orbital periods.

SDSS J1043 was identified as a candidate dwarf nova in the second \textit{GALEX} Ultraviolet Variability catalog (GUVV-2; \citealt{2008AJ....136..259W}). This catalog consists of objects identified as varying by more than 0.6 mag in multiple NUV observations of $\sim$161 deg$^{2}$ of the sky. The repeat observations of SDSS J1043 show two outbursts of $>$2 mag during the two year window. The Catalina Real-Time Transient Survey (CRTS; \citealt{2009ApJ...696..870D}) also shows three outbursts of $>$2 mag during a $\sim$4.5 year time span.

The quiescent spectrum of SDSS J1043 shows the characteristic helium emission of AM CVn binaries, but is unusual in that it shows absorption wings around these lines. Similar spectra are shown by V406 Hya ($P_{\rmn{orb}} = 33.8$ minutes; \citealt{2006MNRAS.365.1109R}) and SDSS J1240 ($P_{\rmn{orb}} = 37.4$ minutes; \citealt{2005MNRAS.361..487R}), in which the DB absorption features are assumed to be due to the accretor (\citealt{2004RMxAC..20..254R}; which is still hot compared to the longer period AM CVn binaries that show no absorption features).

Table \ref{t:ewJ1730} lists the equivalent widths of the main lines seen in the spectra of SDSS J1043 and SDSS J1730.
\begin{table*}
\begin{minipage}{100mm}
\centering
\caption{Equivalent width and Gaussian FWHM for the prominent emission lines in SDSS J1043 and SDSS J1730. Estimated errors are largely due to uncertainty in the continuum.}
\label{t:ewJ1730}
\begin{tabular}{l r r r r}
\hline
Line			& \multicolumn{2}{c}{SDSS J1043} & \multicolumn{2}{c}{SDSS J1730} \\
			& EW (\AA)	& FWHM (km s$^{-1}$) & EW (\AA)	& FWHM (km s$^{-1}$)\footnotetext{Values marked `--' could not be measured reliably.} \\
\hline
He~\textsc{i} 4387	& --		& --			& $-$9 $\pm$ 1	& 1100 $\pm$ 100 \\
He~\textsc{i} 4471	& $-$4 $\pm$ 1	& 3300 $\pm$ 400	& $-$20 $\pm$ 1	& 1100 $\pm$ 100 \\
He~\textsc{ii} 4686	& $-$12 $\pm$ 2	& 5000 $\pm$ 500	& $-$21 $\pm$ 1	& 2600 $\pm$ 200 \\
 \ + He~\textsc{i} 4713	& 		& 			& 		&  \\
He~\textsc{i} 4921	& --		& --			& $-$13 $\pm$ 1	& 820 $\pm$ 20	\\
He~\textsc{i} 5015	& $-$10 $\pm$ 2	& 3200 $\pm$ 500	& $-$19 $\pm$ 1	& 1010 $\pm$ 30	\\
He~\textsc{i} 5875	& $-$16 $\pm$ 2	& 2900 $\pm$ 200	& $-$53 $\pm$ 2	& 820 $\pm$ 10	\\
He~\textsc{i} 6678	& $-$10 $\pm$ 2	& 3000 $\pm$ 300	& $-$45 $\pm$ 2	& 730 $\pm$ 10	\\
He~\textsc{i} 7065	& $-$20 $\pm$ 4	& 3500 $\pm$ 300	& $-$40 $\pm$ 2	& 630 $\pm$ 40	\\
\hline
\vspace{-1cm}
\end{tabular}
\end{minipage}
\end{table*}
The strong helium lines in SDSS J1730 are comparable to those shown by SDSS J1411 ($P_{\rmn{orb}} = 46$ minutes; \citealt{2007PhDT..........R}) and GP Com ($P_{\rmn{orb}} = 46.5$ minutes; \citealt{1981ApJ...244..269N}). Together with the absence of helium absorption at shorter wavelengths, this suggests that SDSS J1730 may be at the long period end of the AM CVn period distribution ($P_{\rmn{orb}} >\sim$ 40 minutes, \citealt{2010ApJ...708..456R}). Whilst the equivalent widths of the lines are similar to those shown by other known AM CVn binaries, the lines are narrow compared with those shown by other systems in our sample. The small FWHM of the emission lines in SDSS J1730 may indicate a low inclination disc. The lack of obvious double peaked structure to the lines also suggests a low inclination for this system \citep{1986MNRAS.218..761H}.

\subsection{AM CVn equivalent width -- period relation}

\begin{figure}
 \includegraphics[width=0.48\textwidth]{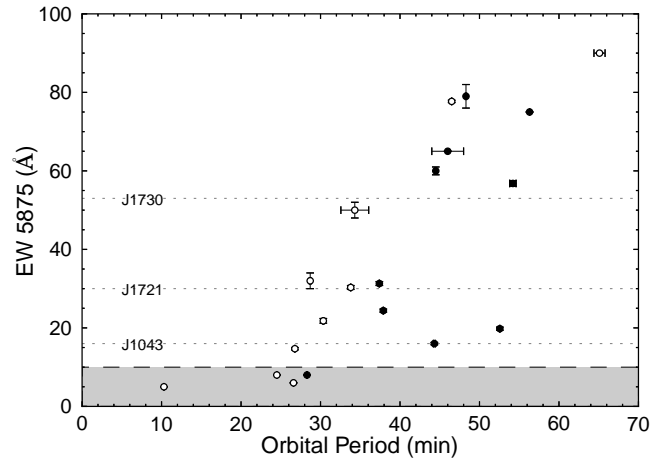}
 \caption{Equivalent width of the He~\textsc{i} 5875 line versus orbital period for AM CVn binaries. The black circles represent those AM CVn binaries discovered from the SDSS. The dotted lines indicate He~\textsc{i} 5875 EWs for the AM CVn binaries discovered via this survey whose periods are unknown. The dashed line and shaded area indicate our detection threshold. \label{f:ewP}}
\end{figure}

The orbital periods and He~\textsc{i} 5875 equivalent widths for AM CVn binaries with emission line spectra are collected in Table \ref{t:P_ew_data}, objects for which no spectrum was available and no measured equivalent width could be found are excluded. Fig. \ref{f:ewP} shows the correlation of the equivalent width of the He~\textsc{i} 5875 emission line with orbital period for AM CVn binaries. There is a general trend for larger equivalent widths with longer binary periods. This is probably due to the drop in continuum flux, against which the lines are measured, due to the cooling of the accreting white dwarf, and drop in accretion continuum as the mass transfer rate decreases towards longer periods. The expected increase in disc size at longer periods, as the orbit expands, may also contribute to the effect. 
\begin{table*}
\begin{minipage}{135mm}
 \centering
 \caption{Orbital period and equivalent width of the He~\textsc{i} 5875 line for the AM CVn binaries with quiescent spectroscopy.}
 \label{t:P_ew_data}
 \begin{tabular}{l r r l}
  \hline
  Object & Orbital period (min) & EW ($-$\AA{}) & References \\
  \hline
  ES Cet & 10.3377 $\pm$ 0.0003 & 5 & \citet{2002PASP..114..129W,2005PASP..117..189E} \\
  CR Boo & 24.5217 $\pm$ 0.0002 & 8 & \citet{1997ApJ...480..383P,2005PASP..117..189E} \\
  V803 Cen & 26.61 $\pm$ 0.02 & 6 & \citet{2007MNRAS.379..176R,2005PASP..117..189E} \\
  PTF J0719 & 26.77 $\pm$ 0.02 & 14.7 $\pm$ 0.3 & \citet{2011ApJ...739...68L} \\
  SDSS J0926 & 28.31 $\pm$ 0.01 & 8 $\pm$ 1 & \citet{2005AJ....130.2230A}, this paper \\
  CP Eri & 28.7 $\pm$ 0.1 & 32 $\pm$ 2 & \citet{1992ApJ...399..680A,2001ApJ...558L.123G} \\
  PTF1 J0943 & 30.35 $\pm$ 0.06 & 21.8 $\pm$ 0.6 & \citet{2012MNRAS.........L} \\
  V406 Hya & 33.80 $\pm$ 0.01 & 30.3 $\pm$ 0.5 & \citet{2006MNRAS.365.1109R} \\
  PTF1 J0435 & 34.31 $\pm$ 1.75 & 49.8 $\pm$ 2.0 & \citet{2012MNRAS.........L} \\
  SDSS J1240 & 37.355 $\pm$ 0.002 & 31.3 $\pm$ 0.5 & \citet{2005MNRAS.361..487R,2006MNRAS.365.1109R} \\
  SDSS J0129 & 37.555 $\pm$ 0.003 & 24.4 $\pm$ 0.4 & \citet{2012MNRAS.........K} \\
  SDSS J1525 & 44.32 $\pm$ 0.12 & 16.0 $\pm$ 0.3 & \citet{2012MNRAS.........K} \\
  SDSS J0804 & 44.5 $\pm$ 0.1 & 60 $\pm$ 1 & \citet{2009MNRAS.394..367R} \\
  SDSS J1411 & 46 $\pm$ 2 & 65 & \citet{2007PhDT..........R,2005AJ....130.2230A} \\
  GP Com & 46.567 $\pm$ 0.003 & 77.7 $\pm$ 0.3 & \citet{1999MNRAS.304..443M,1991ApJ...366..535M} \\
  SDSS J0902 & 48.31 $\pm$ 0.08 & 79 $\pm$ 3 & \citet{2010ApJ...708..456R} \\
  SDSS J1208 & 52.56 $\pm$ 0.02 & 19.8 $\pm$ 0.4 & \citet{2012MNRAS.........K} \\
  SDSS J1642 & 54.2 $\pm$ 0.4 & 56.8 $\pm$ 0.7 & \citet{2012MNRAS.........K} \\
  SDSS J1552 & 56.272 $\pm$ 0.005 & 75 & \citet{2007MNRAS.382.1643R,2005AJ....130.2230A} \\
  V396 Hya & 65.1 $\pm$ 0.7 & 90 & \citet{2001ApJ...552..679R,2009MNRAS.394..367R} \\
  SDSS J1043 & -- & 16 $\pm$ 2 & This paper \\
  SDSS J1721 & -- & 30 $\pm$ 6 & \citet{2010ApJ...708..456R} \\
  SDSS J1730 & -- & 53 $\pm$ 2 & This paper \\
  \hline
 \end{tabular}
\end{minipage}
\end{table*}

This trend would not be expected to hold below a period of $\sim$20 minutes, where the disc is expected to be in a stable high state. ES Cet is included as it shows helium in emission, although it might not be expected to follow the same trend due to its much shorter orbital period. ES Cet in fact shows very strong He~\textsc{ii} 4686 emission, but the weaker 5875 \AA{} line appears to match the trend of the longer period systems reasonably well.

Whilst there is considerable scatter that prevents an accurate estimate of the orbital period based only on the equivalent width, the relation does suggest a period range in which an object might be expected to be found. Based on Fig. \ref{f:ewP}, SDSS J1043 is expected to have an orbital period between 25 and 45 minutes; SDSS J1721 is expected to have an orbital period in the range 25--55 min; and SDSS J1730 is predicted to have an orbital period between 40 and 55 min.

This trend also demonstrates that AM CVn binaries with orbital periods in excess of 30 min should have emission lines sufficiently strong to be well above our detection threshold. We note that the drop in continuum flux at longer orbital periods is expected to make the longer period systems fainter, and less likely to be found within our $g$-band magnitude cut, or the SDSS spectroscopy. This may introduce a selection bias, only the systems with very strong emission appear sufficiently bright to have been observed, and hence there is the possibility of an undetected population of long period, low equivalent width systems.

\subsection{Colours of the sample}

\begin{figure*}
 \includegraphics[width=1.0\textwidth]{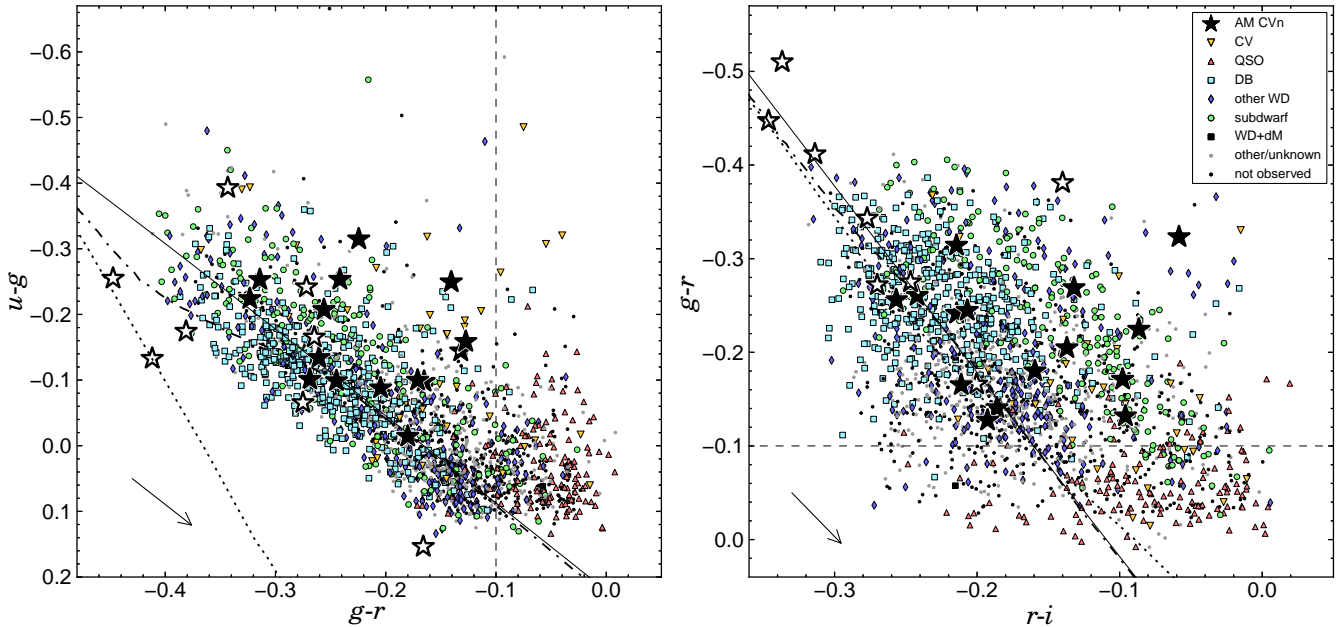}
 \caption{Colours of the known AM CVn binaries (black stars) together with our sample of candidates from SDSS DR7. The filled stars are the confirmed longer period systems, and the AM CVn binaries discovered via their emission line spectra in the SDSS or this survey. The open stars represent those systems to which we do not necessarily expect to be sensitive, from left to right (left hand panel) or from top to bottom (right hand panel), these are HM Cnc (right hand panel only), SDSS J1908, SDSS J2047, PTF1 J2219, ES Cet, PTF1 J0857, PTF1 J0943, PTF1 J0435 (left hand panel only) and CR Boo.
 The orange inverted triangles represent CVs. Quasars, DB white dwarfs, other types of white dwarf, subdwarfs, and WD + M dwarf systems are represented by red triangles, cyan squares, blue diamonds, green circles and black squares respectively. Grey dots indicate those candidates that could not be classified, and black dots those that have not yet been observed.
 The solid line shows the blackbody cooling track, the dotted and dot-dashed lines represent the DA and DB white dwarf cooling tracks, and arrows represent reddening vectors for an extinction $A(g)=0.2$. The dashed line represents the new colour cut.\label{f:colourplots}}
\end{figure*}
The ($u-g$, $g-r$) and ($g-r$, $r-i$) colour diagrams of the sample and the colours of a subset of the known AM CVn binaries (black stars) are shown in Fig. \ref{f:colourplots}. The most significant of the classifications from Table \ref{t:class} are also indicated with different markers. The seven systems discovered in the observed part of the sample further constrain the region of colour space occupied by AM CVn binaries. The colours of AM CVn itself may be affected by saturation due to its brightness, it has therefore been excluded. $u-g$ values calculated from the $UBV$ photometry of CR Boo given in table 1 of \citet{1987ApJ...313..757W} (using conversions given by \citealt{2006A&A...460..339J}), suggest that the colours can vary significantly over an outburst cycle. CR Boo was likely to have been in an intermediate state at the time of the SDSS photometry ($g = 15.6$ compared to $g \simeq 17.1$ in quiescence, based on photometry from \citealt{1987ApJ...313..757W}), which may have caused the unusual $u-g$ colour.

The initial selection of the sample \citep{2009MNRAS.394..367R} was based on the colours of the nine emission line AM CVn binaries with SDSS photometry known at the time. Fig. \ref{f:colourplots} shows the colours of 24 of the 35 currently known AM CVn binaries. Whilst many of the systems discovered since have been found through this survey, all long period systems, and most intermediate period systems with quiescent SDSS photometry, fall within our colour box (see also \citealt{2012MNRAS.........L}, for a discussion of the colours of outbursting systems found by the Palomar Transient Factory). It should be noted that the survey targets the specific area of colour--colour space in which the AM CVn binaries in the SDSS spectroscopic database are found. We are therefore only capable of finding systems that share these colours, and any long period AM CVn binaries that lie outside this colour box will remain undetected. The biases in the SDSS spectroscopic database are discussed by \citet{2007MNRAS.382..685R,2009MNRAS.394..367R}, they conclude that it is unlikely that a large fraction of AM CVn binaries lie outside this region. This more robust knowledge of the location in colour space of AM CVn binaries, and other objects in this region, allows us to improve upon the initial selection criteria and reduce the sample size.

It is clear from Fig. \ref{f:colourplots} that the known AM CVn binaries with SDSS photometry occupy the region with
\[
 g-r < -0.1;
\]
approximately 81 per cent of the quasars (red triangles) have $g-r$ above this limit, and many could be safely removed. It should be noted that the spectroscopic completeness of the SDSS increases with increasing $g-r$ over the densely populated area of this region, where objects have been targeted for quasars (Fig. \ref{f:sloancompleteness}). The original sample of systems from the SDSS spectroscopic database is thus slightly biased towards the red cut-off. This makes it unlikely that a large fraction of AM CVn binaries lie beyond $g-r = -0.1$, and we consider the risk to our completeness to be small in comparison to the gain in efficiency.

\begin{figure*}
 \includegraphics[width=1.0\textwidth]{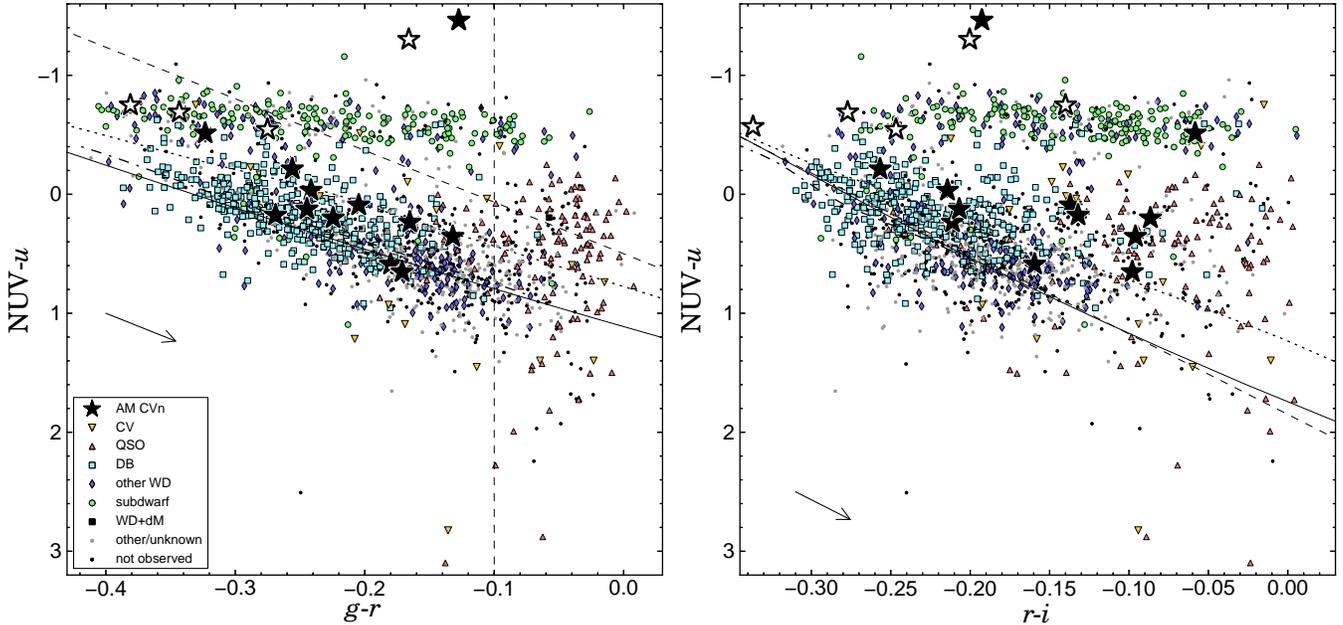}
 \caption{Colours of the known AM CVn binaries (black stars) together with our sample of candidates from SDSS DR7 combining \textit{GALEX} NUV detections. The symbols have the same meanings as in Fig. \ref{f:colourplots}. The open stars represent, from top to bottom, CR Boo, PTF1 J2219, ES Cet, HM Cnc (right hand panel only) and PTF1 J0857. The large negative $\rmn{NUV}-u$ colour of CR Boo and SDSS J1043 is caused by their varying brightness states between the epochs of \textit{GALEX} and SDSS observations. The solid line shows the blackbody cooling track, the dotted and dot-dashed lines represent the DA and DB white dwarf cooling tracks, and arrows represent reddening vectors for an extinction $A(g)=0.2$. A significant portion of DB white dwarfs lie below of slightly above the blackbody cooling track in ($\rmn{NUV}-u$, $r-i$) colour space, whereas the AM CVn binaries all lie above this track. The dashed lines represent the new colour cuts. \label{f:GALEXcolourplots}}
\end{figure*}

The addition of UV photometry allows further examination of the sample in additional colour spaces. Fig. \ref{f:GALEXcolourplots} shows colour diagrams of the part of the sample with \textit{GALEX} NUV detections, these diagrams include 17 of the known AM CVn binaries. UV extinction was estimated using the $R_V=3.1$ prescription of \citet{1989ApJ...345..245C} and the full Galactic reddening according to \citet{1998ApJ...500..525S}, in order to match the procedure used for the SDSS. Note that CR Boo and SDSS J1043 were much closer to their maximum brightness states when observed by \textit{GALEX}, compared with their SDSS photometry, causing the large negative $\rmn{NUV}-u$. There are also several other objects with $\rmn{NUV}-u < -1$, for which different brightness states are a possibility ($\rmn{NUV}-u \simeq -1$ is the limit for a hot blackbody).

The $\rmn{NUV}-u$ colour greatly increases the separation of some object types compared with only Sloan colours. The subdwarfs and white dwarfs occupy mostly separated areas of the diagrams in Fig. \ref{f:GALEXcolourplots}; $\rmn{NUV}-u$ also separates the subdwarfs and the quasars. Many of the subdwarfs (green circles) in the upper right corner of the left hand panel of Fig. \ref{f:GALEXcolourplots} could also be discarded, as the AM CVn binaries do not spread into this region. By requiring
\[
 \rmn{NUV}-u > 4.34 (g-r) + 0.5, 
\]
46 per cent of the subdwarfs can be removed, and the sample size is reduced by 11 per cent.

The DB white dwarfs (cyan squares) lie around and close to the blackbody cooling track in ($\rmn{NUV}-u$, $r-i$) colour space, whereas the AM CVn binaries all lie above. With a cut close to this cooling track, and parallel to the reddening vector,
\[
 \rmn{NUV}-u < 6.76 (r-i) + 1.85,
\]
approximately one fifth of the DB white dwarfs can be removed from the sample, significantly reducing the numbers of the main contaminant, and reducing the total size of the sample to $\sim$1500.

We emphasize that these additional cuts increase the risk of missing some AM CVn binaries that are present in the SDSS photometry, making it more difficult to judge our completeness, and the accuracy of our space density. However, the increase in efficiency offers a significant advantage in reducing the observing time required to complete such a survey.

\begin{figure*}
 \includegraphics[width=1.0\textwidth]{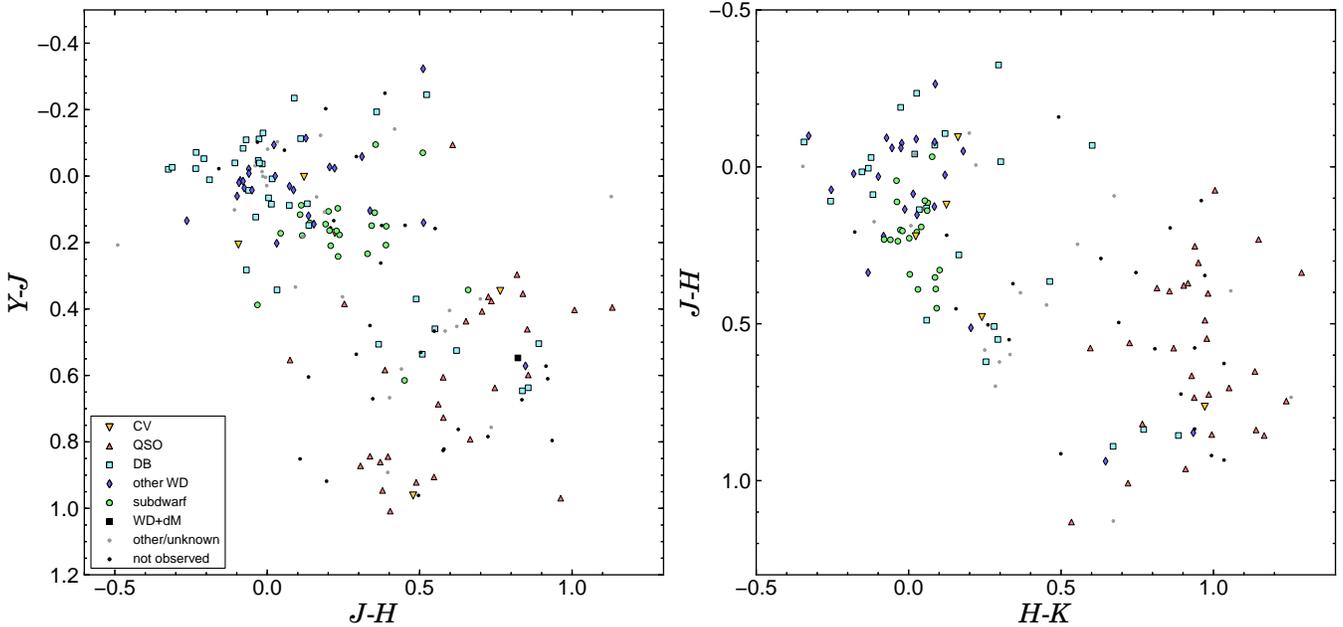}
 \caption{Infrared colours of our sample. The symbols have the same meanings as in Fig. \ref{f:colourplots}. Both diagrams show the potential for removing quasars that infrared data provides, however, greater coverage is required to enable reliable cuts to be made. \label{fig:UKIDSScolourplots}} 
\end{figure*}
The UKIDSS colours of the sample are shown in Fig. \ref{fig:UKIDSScolourplots}. The separation of quasars from the rest of the sample is clear even with the limited sky coverage (see also \citealt{2009AJ....137.3884R,2010MNRAS.406.1583W}; for a discussion of quasar selection via infrared colours). As only four of the known AM CVn binaries have been covered by UKIDSS, and none with sufficient detections whilst in the same brightness state to appear in Fig. \ref{fig:UKIDSScolourplots}, any attempt to remove targets from the sample based on infrared colours would be unreliable. Greater coverage with infrared photometry could, however, allow further improvements to the sample size. The \textit{WISE} mission \citep{2010AJ....140.1868W} has recently conducted an all sky infrared survey, however, the depth of the survey is insufficient for the majority of known AM CVn binaries.

\subsection{The AM CVn binary space density}

The discovery of only seven AM CVn binaries in the 72 per cent of the sample that has been observed, taking into account our spectroscopic completeness (Fig. \ref{f:completeness}), indicates the presence of significantly fewer AM CVn binaries in the SDSS photometric database than initially estimated by \citet{2007MNRAS.382..685R,2009MNRAS.394..367R} ($\sim$40 new AM CVn binaries in the original sample of $\sim$1500 candidates from SDSS DR6; there are $\sim$2000 objects in the DR7 sample). This suggests that the space density of AM CVn binaries is lower than previously predicted, or that there is some problem with our understanding of the biases in our sample or the population models. Based on the analysis presented in \citet{2007MNRAS.382..685R,2009MNRAS.394..367R}, we consider the former to be likely.

Fig. \ref{f:completeness} shows that our follow-up is almost complete to a depth of $g$ = 19. We use this essentially complete sample of the brighter objects in the AM CVn colour box to adjust earlier space density estimates, although the large uncertainty resulting from such small samples must be noted. Following the prescription of \citet{2007MNRAS.382..685R}, we calculate the expected magnitude distribution of the AM CVn binaries in the SDSS photometric database. The number of systems expected is then compared to the number found in the SDSS DR7 area.

Based on the optimistic model from \citet{2001A&A...368..939N,2004MNRAS.349..181N}, and the six AM CVn binaries discovered in the SDSS spectroscopy, \citet{2007MNRAS.382..685R} calculated an observed space density 1.5~$\times$~10$^{-6}$~pc$^{-3}$. This corresponds to an expected 35 AM CVn binaries to a depth of $g$ = 20.5 (see Table \ref{t:spacedensity}), in the DR5 photometric database (note that this number is essentially independent of the formation channel; see table 1 of \citealt{2007MNRAS.382..685R}). To find the distribution of systems the population synthesis predicts for our sample, we multiply the \citet{2007MNRAS.382..685R} distribution by the ratio of the photometric area of DR7 (11663 deg$^2$) to that of DR5 (8000 deg$^2$), $\sim$1.46. This gives expected numbers of 11 systems with $g \le$~19, and 51 systems with $g \le$~20.5.
\begin{table*}
\begin{minipage}{132mm}
\centering
\caption{Expected numbers of AM CVn binaries in the AM CVn colour box and corresponding space densities. Note that \citealt{2007MNRAS.382..685R} use a depth of $g =$~21, whereas our survey has a limit of $g =$~20.5.}
\label{t:spacedensity}
\begin{tabular}{l r r r r}
\hline
Distribution				& \multicolumn{3}{c}{Expected number of AM CVn binaries} & Space density (pc$^{-3}$) \\
					& $g \le$ 19	& $g \le$ 20.5	& $g \le$ 21	& \\
\hline
\citet{2007MNRAS.382..685R} DR5 area	& 8		& 35		& 53		& 1.5~$\times$~10$^{-6}$ \\
\citet{2007MNRAS.382..685R} DR7 area	& 11		& 51		& 77		& 1.5~$\times$~10$^{-6}$ \\
Scaled to $g \le$ 19 sample (DR7)	& 4		& 18 		& 27		& (5 $\pm$ 3)~$\times$~10$^{-7}$ \\
\hline
\end{tabular}
\end{minipage}
\end{table*}

We use the 4 long period emission line AM CVn binaries found in our essentially complete $g \le$ 19 sample to scale the expected numbers and observed space density (Table \ref{t:spacedensity}). This gives the total number of AM CVn binaries expected in the DR7 photometry ($g \le$ 20.5) as 18, corresponding to a space density for AM CVn binaries of 5~$\times$~10$^{-7}$~pc$^{-3}$.

The small sample of 4 known AM CVn binaries with $g \le$ 19 contributes an intrinsic uncertainty of 50 per cent to the derived space density. \citet{2007MNRAS.382..685R} estimate the uncertainties in the parametrization of temperature and absolute magnitude with orbital period as leading to $\sim$10 per cent and $\sim$32 per cent variations in the result. Combining these contributions leads to an estimated 60 per cent uncertainty in our value for the space density.

The space density calculations assume models for the intrinsic magnitude distribution of the AM CVn binaries \citep{2007MNRAS.382..685R}. It is clear from Fig. \ref{f:completeness} that we find an increasing number of these systems towards fainter magnitudes. This may reduce the discrepancy between the previously predicted space density and the numbers of systems found so far in our sample, but it may also indicate a significant deviation from the modelled magnitude distribution.

\section{Discussion}

The results from our survey so far significantly improve our knowledge of the colours of AM CVn binaries, and the other objects found in the same region of colour space. The seven new AM CVn binaries lie in the central area of the selected colour region, implying that the initial selection should indeed have contained the majority of AM CVn binaries, but also that it could have been more efficient. The reddest part of the $g-r$ distribution is largely made up of quasars and unclassified objects, and can be removed from the sample.

Cross matching the sample with the \textit{GALEX} catalogue provides a further colour space that proves to be very useful in separating different classes of object. The majority of the subdwarfs in our sample are brighter in NUV relative to $u$ than the white dwarfs, causing the gap between the two populations that is not seen with only Sloan colours. That the AM CVn binaries are slightly redder than many of the DB white dwarfs in $r-i$, perhaps due to the contribution of the accretion disc, is very useful when combined with their distribution in $\rmn{NUV}-u$. This increased splitting of DB white dwarfs from the other objects in the sample in ($\rmn{NUV}-u$, $r-i$) colour space, allows about one fifth of this major contaminant to be safely discarded. Subdwarfs and AM CVn binaries become increasingly separated with increasing $g-r$ in ($\rmn{NUV}-u$, $g-r$) colour space, such that the majority of the redder subdwarfs can be cut. Making these cuts parallel to the reddening vectors ensures that objects are not removed unintentionally due to poorly estimated extinction. The small number of objects with $\rmn{NUV}-u < -1$ or $\rmn{NUV}-u > 1.5$ are kept, as the large values may be caused by varying brightness state between the epochs of \textit{GALEX} and SDSS observations.

Combining the two new cuts involving $\rmn{NUV}-u$, and the more efficient $g-r$ cut-off, in addition to the original $u-g$, $g-r$ and $r-i$ colour cuts used to produce the sample, reduces the total size by 43 per cent. Removing 268 of the 544 remaining targets, brings the goal of detecting all of the hidden AM CVn population much closer to completion. These extra cuts, whilst greatly increasing our efficiency, risk reducing our completeness. Although the new colour criteria avoid the regions in which AM CVn binaries have been found, it is possible that there exists some as yet undetected part of the population that deviates from the current distribution. Any such systems would not be expected to represent the majority of the population, and we consider the risk in losing them to be small. Increasing the efficiency of our survey becomes increasingly important as we move to fainter targets, where greater numbers of AM CVn binaries have been found (see Fig. \ref{f:completeness}), in order to keep the project feasible.

\begin{figure}
 \centering
 \includegraphics[width=0.48\textwidth]{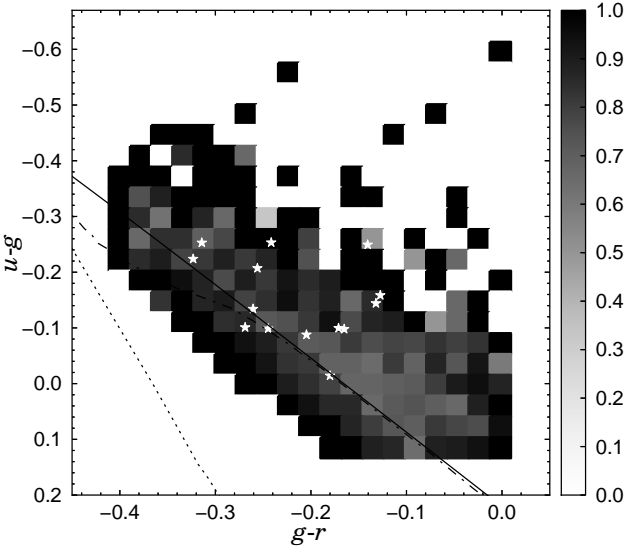}
 \caption{Completeness of the survey as a function of colour, $u-g$ and $g-r$, to a limiting magnitude $g=20.5$ (dereddened), including both SDSS spectroscopy and our own. The AM CVn binaries with spectra taken as part of the SDSS or this survey, are indicated by star symbols. The solid line marks the blackbody cooling track, the dotted and dot-dashed lines indicate model cooling sequences for DA and DB white dwarfs. \label{f:colcomp}} 
\end{figure}
Fig. \ref{f:colcomp} shows the spectroscopic completeness of the survey colour box, combining both SDSS spectroscopy and our own. Our spectroscopic followup has been conducted without any specific colour bias beyond the initial selection criteria. However, the magnitude bias shown in Fig. \ref{f:completeness} leads to a slight bias towards the bluer part of the ($u-g$, $g-r$) distribution. The lower density of objects at the edges of our survey area leads to a significantly higher completeness. There is no clear bias that would affect the AM CVn distribution we observe. 

The observed trend of increasing equivalent width of emission lines with orbital period, coupled with the expectation that AM CVn binaries will evolve rapidly to longer periods, allows us to be confident that we are able to detect the majority of the population. This is important as we intend to use the number we find to estimate the space density of the entire population. The equivalent widths of the SDSS AM CVn binaries with unknown periods suggest that many of them are at the shorter period end of the distribution that we would expect (see \citealt{2007MNRAS.382..685R}). This suggests that there should be more systems with $P_{\rmn{orb}} > 50$ min yet to be discovered in our sample, as this is where AM CVn binaries should accumulate \citep{2001A&A...368..939N,2012ApJ...758..131N}.

Using our essentially complete $g \le$~19 sample, we calculate the expected number of AM CVn binaries in SDSS DR7 to be 18 $\pm$ 9 ($g \le$~20.5). Since we know of 15 longer period systems with SDSS photometry\footnote{This number includes the 6 systems discovered in the SDSS spectroscopy, the 7 systems discovered via our survey, GP Com, and SDSS J1427-0123 -- which has not been observed spectroscopically by us or the SDSS.}, this suggests that there should be 3 undiscovered systems left in our sample, though the considerable uncertainty on this number leaves us unable to make firm predictions. From the spectroscopic completeness as a function of $g$-band magnitude (Fig. \ref{f:completeness}), we would expect to find $\sim$5 more systems in the remainder of our sample, this is consistent with the number derived from our revised space density estimate, but also has a large uncertainty.

\citet{2007MNRAS.382..685R} calculated the AM CVn space density to be 1.5~$\times$~10$^{-6}$~pc$^{-3}$, based on the six systems found in the SDSS spectroscopic database. Our revised value is 3 times lower than this, and 50 times lower than the optimistic model from the \citet{2001A&A...368..939N} population synthesis predicts.

\textcolor{black}{\citet{2012ApJ...758..131N} suggest that the lower space density could be explained by a change to the galactic disc model used by \citet{2001A&A...368..939N}. The distances of AM CVn binaries from our solar system, and hence the amplitude of their gravitational wave signal, depends on the assumed scale height for AM CVn binaries in the Galaxy. In the proposed alteration \citep{2012ApJ...758..131N}, the old systems, that are the ones we would detect as emission line systems in the SDSS sample, have a larger scaleheight, and so are farther away. However, the young systems, that are the majority of those that would be detected via their gravitational wave emission, remain unchanged from the original distribution \citep{2001A&A...368..939N,2004MNRAS.349..181N}. This results in the same total numbers of AM CVn binaries in the Galaxy, but the long period systems are at greater distances, and are hence fainter. This could explain why optical surveys find less systems than predicted by \citet{2001A&A...368..939N} and \citet{2007MNRAS.382..685R}, but has no significant effect on the numbers of AM CVn binaries detectable by space-based gravitational wave missions (compare cases 1 and 5 shown in table 2 and table 3 of \citealt{2012ApJ...758..131N}).}

Alternatively, the lack of systems could be explained as problems with the binary population synthesis; if fewer systems than expected survive to become stable mass transferring AM CVn binaries, the predicted space density would fall. It should be noted that there is a large intrinsic uncertainty in the population synthesis numbers, this is why \citet{2001A&A...368..939N} have optimistic and pessimistic models. It is also possible that the discrepancy is so large that it can only be explained by changes to both the population synthesis and the expected brightness distribution of the AM CVn binaries.

This survey is based on the colours of the known emission line AM CVn binaries, and hence targets the long period systems ($P_{\rmn{orb}} > 30$ min). Whilst the majority of all systems fall within our colour box, we are only sensitive to intermediate period systems in their low state. In their high state these AM CVn binaries normally appear much like DB white dwarfs, and if they were also in their high state at the time the SDSS observed them, we would be unable to recognise them as AM CVn binaries. It should be noted that a small percentage of the total AM CVn population should be found as outbursting systems. We estimate that AM CVn binaries spend less than 6 per cent of their lifetime in this state. These frequent outbursting systems ($20 < P_{\rmn{orb}} < \sim30$ min) can be detected more efficiently from their variability, using synoptic surveys, as in the case of PTF1 J0719+4858 \citep{2011ApJ...739...68L,2012MNRAS.........L}. Follow-up of these systems, however, may be more challenging as they are only expected to be found in significant numbers by increasing the depth of the search. SDSS J1043 is particularly interesting as it represents the overlap between these two methods, detected both as a result of its colours and its variability. These complimentary methods mostly sample different parts of the AM CVn orbital period distribution, and together, will lead to a better understanding of both this period distribution, and the space density.

\section{Conclusion}

We have presented an update on the status of our spectroscopic survey aimed at uncovering the expected hidden population of AM CVn binaries in the SDSS photometric database. The results currently indicate a lower space density than predictions suggest. Based on the brighter part of our sample, we calculate an observed space density of (5~$\pm$~3)~$\times$~10$^{-7}$~pc$^{-3}$.

We have reported the discovery of two candidate AM CVn binaries found via this survey, SDSS J104325.08+563258.1 and SDSS J173047.59+554518.5. SDSS J1043 exhibits the helium absorption and low equivalent width emission lines shown by the AM CVn binaries with orbital periods below $\sim$40 minutes. SDSS J1730 shows strong helium emission lines with no helium absorption at shorter wavelengths; together with the large He~\textsc{i} 5875 equivalent width, this suggests it has an orbital period longer than $\sim$40 minutes.

As we push towards fainter targets, removing candidates that are very unlikely to be AM CVn binaries is important in order to ensure that observing the remainder of the sample remains practically achievable. Combining new constraints provided by \textit{GALEX} fluxes with our increased knowledge of the region of colour space occupied by AM CVn binaries, leads to a 43 per cent reduction in the size of the sample. This cuts the number of objects still requiring observation by 49 per cent, to 275. This should result in a corresponding increase in our AM CVn hit-rate, and allow the remaining long period AM CVn binaries hidden in the SDSS photometric database to be uncovered. This will enable the estimates of the space density to be greatly refined, and should reduce the several orders of magnitude uncertainty in its value. Establishing this will provide constraints for common envelope evolution models, allow more accurate modelling of the gravitational wave signal and foreground, and lead to a better understanding of the stability of mass transfer in possible progenitor systems.

\section*{Acknowledgements}

We thank the anonymous referee for useful comments and suggestions.
TRM, DS and CMC acknowledge support from the Science and Technology Facilities Council (STFC) grant no. ST/F002599/1.
Funding for the SDSS and SDSS-II has been provided by the Alfred P. Sloan Foundation, the Participating Institutions, the National Science Foundation, the U.S. Department of Energy, the National Aeronautics and Space Administration, the Japanese Monbukagakusho, the Max Planck Society, and the Higher Education Funding Council for England. The SDSS Web Site is http://www.sdss.org/.
\textit{GALEX} is a NASA Small Explorer. The mission was developed in cooperation with the Centre National d’Etudes Spatiales of France and the Korean Ministry of Science and Technology. This research has made use of the SIMBAD database, operated at CDS, Strasbourg, France; and NASA's Astrophysics Data System Bibliographic Services.
Several figures make use of P. Bergeron's synthetic white dwarf colours, taken from http://www.astro.umontreal.ca/$\sim$bergeron/CoolingModels.
Fig. \ref{f:GALEXcolourplots} makes use of D. Koester's white dwarf atmosphere models.

\appendix

\section{Target catalogue}

\label{A:candtable}
\textcolor{black}{Our target catalogue of AM CVn candidates selected from SDSS DR7, with \textit{GALEX} UV and Sloan magnitudes, and classifications based on our spectroscopic followup, is given in Table \ref{t:catalogue}. The $g$-band extinction assuming full Galactic reddening according to \citet{1998ApJ...500..525S} is also given. It is assumed that at the galactic latitudes of the SDSS, the majority of AM CVn binaries will be further from the plane than the scaleheight of dust in the Galaxy, see \citet{2009MNRAS.394..367R}.}

\landscape
\begin{table}
 \centering
 \caption{Target catalogue with \textit{GALEX} UV magnitudes and classifications. Suffixed colons indicate an uncertain classification. Magnitudes given use the AB photometric system. This is a sample of the full table, which is available in the online version of this article (see Supporting Information). \textcolor{black}{The full spectroscopic catalogue is available in electronic form at the CDS
(http://cdsweb.u-strasbg.fr/).}}
\label{t:catalogue}
\begin{tabular}[c]{l r c c c c c c c c c c c c c c c l}
\hline
RA (J2000) & Dec (J2000) & FUV & $\sigma_\rmn{FUV}$ & NUV & $\sigma_\rmn{NUV}$ & $u$ & $\sigma_u$ & $g$ & $\sigma_g$ & $r$ & $\sigma_r$ & $i$ & $\sigma_i$ & $z$ & $\sigma_z$ & $A(g)$ & Classification \\
\hline
00:00:48.78	& $-$09:12:09.1	& --	& --	& 21.05	& 0.26	& 20.19	& 0.06	& 20.14	& 0.02	& 20.25	& 0.03	& 20.27	& 0.05	& 21.31	& 0.43	& 0.14	&  \\
00:01:00.76	& $+$24:04:18.3	& --	& --	& --	& --	& 20.41	& 0.05	& 20.36	& 0.02	& 20.42	& 0.03	& 20.59	& 0.04	& 20.57	& 0.15	& 0.36	&  \\
00:01:06.22	& $+$25:03:30.1	& 21.15	& 0.21	& --	& --	& 19.49	& 0.04	& 19.50	& 0.01	& 19.67	& 0.02	& 19.90	& 0.03	& 20.19	& 0.16	& 0.27	& DB \\
00:01:11.66	& $+$00:03:42.6	& 18.44	& 0.08	& 18.88	& 0.07	& 19.21	& 0.03	& 19.24	& 0.01	& 19.31	& 0.02	& 19.39	& 0.02	& 19.46	& 0.10	& 0.11	& DA: \\
00:01:51.26	& $+$32:56:58.7	& 15.05	& 0.02	& 15.48	& 0.01	& 16.05	& 0.01	& 16.19	& 0.00	& 16.41	& 0.00	& 16.52	& 0.00	& 16.61	& 0.01	& 0.18	& sdB: \\
00:02:05.57	& $+$00:20:41.7	& 20.15	& 0.18	& 20.32	& 0.18	& 19.96	& 0.05	& 20.09	& 0.02	& 20.37	& 0.03	& 20.55	& 0.05	& 20.86	& 0.28	& 0.12	& DB \\
00:02:13.15	& $-$01:22:13.8	& 19.20	& 0.12	& 19.61	& 0.11	& 19.96	& 0.06	& 20.08	& 0.02	& 20.23	& 0.03	& 20.31	& 0.05	& 20.17	& 0.17	& 0.15	&  \\
00:02:19.66	& $+$15:14:31.6	& --	& --	& 20.52	& 0.10	& 20.10	& 0.05	& 20.09	& 0.02	& 20.26	& 0.02	& 20.51	& 0.04	& 20.94	& 0.29	& 0.17	& DAQ: \\
00:02:23.06	& $+$27:23:58.5	& 17.88	& 0.08	& 17.41	& 0.02	& 17.44	& 0.01	& 17.60	& 0.01	& 17.85	& 0.01	& 18.10	& 0.01	& 18.34	& 0.03	& 0.17	& DB \\
00:03:02.98	& $-$03:24:29.7	& --	& --	& 21.28	& 0.28	& 20.43	& 0.08	& 20.35	& 0.02	& 20.36	& 0.03	& 20.49	& 0.05	& 20.03	& 0.16	& 0.15	&  \\
00:04:26.95	& $+$24:32:58.9	& 21.75	& 0.28	& 21.04	& 0.10	& 19.98	& 0.05	& 19.91	& 0.02	& 20.00	& 0.02	& 20.20	& 0.03	& 20.29	& 0.14	& 0.36	& DBZ: \\
00:07:29.31	& $+$06:19:16.7	& 18.18	& 0.07	& 17.96	& 0.04	& 17.74	& 0.01	& 17.72	& 0.01	& 17.69	& 0.01	& 17.70	& 0.01	& 17.03	& 0.01	& 0.29	& QSO \\
00:08:16.25	& $+$15:46:09.6	& --	& --	& --	& --	& 18.70	& 0.02	& 18.69	& 0.01	& 18.85	& 0.01	& 19.00	& 0.02	& 19.22	& 0.06	& 0.28	& DB \\
00:08:47.25	& $-$05:12:17.1	& 16.37	& 0.02	& 16.74	& 0.01	& 17.27	& 0.01	& 17.38	& 0.00	& 17.64	& 0.01	& 17.80	& 0.01	& 17.89	& 0.02	& 0.11	& sdB: \\
00:13:31.44	& $-$09:52:52.5	& --	& --	& --	& --	& 20.01	& 0.06	& 20.09	& 0.02	& 20.26	& 0.03	& 20.48	& 0.06	& 20.56	& 0.29	& 0.14	&  \\
23:49:34.16	& $+$34:58:15.9	& 20.48	& 0.23	& 19.57	& 0.08	& 19.00	& 0.02	& 18.97	& 0.01	& 19.07	& 0.01	& 19.22	& 0.02	& 19.35	& 0.07	& 0.20	& DB: \\
23:49:53.93	& $+$34:18:55.4	& --	& --	& 21.10	& 0.11	& 20.50	& 0.05	& 20.46	& 0.02	& 20.53	& 0.03	& 20.67	& 0.04	& 21.00	& 0.19	& 0.24	&  \\
23:50:02.23	& $+$35:00:21.7	& 22.35	& 0.43	& 20.86	& 0.17	& 20.27	& 0.06	& 20.37	& 0.03	& 20.55	& 0.04	& 20.64	& 0.06	& 20.92	& 0.26	& 0.20	& DB: \\
23:50:28.70	& $+$35:50:33.3	& --	& --	& 21.20	& 0.23	& 20.46	& 0.06	& 20.42	& 0.02	& 20.47	& 0.03	& 20.68	& 0.04	& 20.80	& 0.16	& 0.30	&  \\
23:51:25.70	& $+$06:33:05.4	& --	& --	& 20.37	& 0.17	& 19.20	& 0.03	& 19.05	& 0.01	& 19.08	& 0.01	& 19.21	& 0.02	& 19.33	& 0.06	& 0.32	&  \\
23:52:19.59	& $+$32:17:08.9	& 20.84	& 0.21	& 20.48	& 0.13	& 20.18	& 0.04	& 20.26	& 0.02	& 20.50	& 0.03	& 20.70	& 0.04	& 20.75	& 0.15	& 0.18	& DB \\
23:54:48.96	& $+$31:31:59.6	& --	& --	& 20.54	& 0.21	& 19.36	& 0.03	& 19.38	& 0.01	& 19.53	& 0.02	& 19.72	& 0.03	& 19.92	& 0.10	& 0.18	&  \\
23:55:11.28	& $+$06:48:22.6	& --	& --	& --	& --	& 18.65	& 0.02	& 18.53	& 0.01	& 18.51	& 0.01	& 18.52	& 0.01	& 18.37	& 0.02	& 0.20	& QSO \\
23:55:44.79	& $+$29:46:19.4	& --	& --	& --	& --	& 19.24	& 0.02	& 19.08	& 0.01	& 19.15	& 0.01	& 19.27	& 0.02	& 19.50	& 0.05	& 0.22	& DZ \\
23:55:49.59	& $+$36:38:48.5	& --	& --	& 21.53	& 0.11	& 20.53	& 0.05	& 20.41	& 0.02	& 20.42	& 0.02	& 20.51	& 0.04	& 20.66	& 0.14	& 0.46	&  \\
23:57:14.91	& $+$30:07:03.9	& --	& --	& --	& --	& 19.45	& 0.03	& 19.48	& 0.01	& 19.67	& 0.02	& 19.85	& 0.02	& 20.08	& 0.08	& 0.18	&  \\
23:58:44.16	& $-$10:22:02.3	& 19.15	& 0.11	& 18.89	& 0.07	& 19.02	& 0.03	& 19.13	& 0.01	& 19.42	& 0.02	& 19.62	& 0.03	& 20.04	& 0.16	& 0.13	& DB \\
23:58:58.88	& $+$32:40:19.6	& --	& --	& --	& --	& 17.67	& 0.01	& 17.63	& 0.01	& 17.65	& 0.01	& 17.67	& 0.01	& 16.99	& 0.01	& 0.17	& QSO \\
23:59:06.73	& $-$11:12:32.1	& 18.37	& 0.08	& 18.59	& 0.06	& 19.34	& 0.03	& 19.66	& 0.02	& 19.92	& 0.03	& 20.11	& 0.04	& 20.41	& 0.21	& 0.13	& sdOB: \\
23:59:45.79	& $+$25:19:09.5	& 18.25	& 0.05	& 17.81	& 0.02	& 17.50	& 0.01	& 17.35	& 0.00	& 17.40	& 0.01	& 17.43	& 0.01	& 17.01	& 0.01	& 0.20	& QSO \\
\hline
\end{tabular}
\end{table}
\label{lastpage}
\endlandscape

%
%

\end{document}